\documentclass[a4paper,twocolumn,nofootinbib]{revtex4}

\usepackage[top=2.8cm, bottom=2.8cm, left=2.4cm, right=2.4cm]{geometry}
\usepackage[utf8]{inputenc}  
\usepackage[T1]{fontenc}     
\usepackage{amsmath,amssymb,lmodern} 
\usepackage{amsbsy}
\usepackage{graphicx}
\usepackage{hyperref}
\usepackage{lipsum}

\begin{document}

\title{Bosonic structure of realistic SO(10) SUSY cosmic strings}

\author{Erwan Allys}
\email{allys@iap.fr}
\affiliation{$\mathcal{G}\mathbb{R}\epsilon\mathbb{C}\mathcal{O}$, Institut d’Astrophysique de Paris, UMR 7095, 98 bis bd
  Arago, 75014 Paris, France \\ UPMC Universit\'e Paris 6 et CNRS,
  Sorbonne Universit\'es}

\begin{abstract}
We study the bosonic structure of $F$-term Nambu-Goto cosmic strings forming in a realistic SO(10) implementation, assuming standard hybrid inflation. We describe the supersymmetric grand unified theory, and its spontaneous symmetry breaking scheme in parallel with the inflationary process. We also write the explicit tensor formulation of its scalar sector, focusing on the sub-representations singlet under the standard model, which is sufficient to describe the string structure. We then introduce an ansatz for abelian cosmic strings, discussing in details the hypothesis, and write down the field equations and boundary conditions. Finally, after doing a perturbative study of the model, we present and discuss the results obtained with numerical solutions of the string structure.
\end{abstract}

\date{\today}

\maketitle

%%%%%%%%%%%%%%%%%%%%%%%%%%%%%%%%%%%%%%%%%%%%%%%%%%%%%%%%%%%%%%%%%
%%%%%%%%%%%%%%%%%%%%%%%%%%%%%%%%%%%%%%%%%%%%%%%%%%%%%%%%%%%%%%%%%
\section{Introduction}

The phenomenology of the early universe gives access to high energy physics models, such as Grand Unified Theories (GUTs). Indeed, we know that the Spontaneous Symmetry Breaking (SSB) down to the Standard Model (SM) of such theories around $10^{16}$GeV must produce topological defects, e.g. cosmic strings \cite{Kibble:1980mv,Hindmarsh:1994re}. The observation of such objects, for instance in the Cosmic Microwave Background (CMB) \cite{Bouchet:2000hd,Bevis:2007gh,Ringeval:2010ca,Ade:2013xla,Brandenberger:2013tr}, allows then to put constraints on the string energy per unit length, and thus on the GUT which led their formation. In the past, some work has already been done to study the structures and properties of such strings, see e.g. \cite{Kibble:1982ae,Ma:1992ky,Davis:1996sp,Davis:1997bs,Morris:1997ua,Davis:1997ny}.

In order to have a complete understanding of these objects, it is necessary to study them in a realistic GUT context, and not only through toy models which contain only the minimal field content necessary to describe this kind of defects. Such a work was done in Ref.~\cite{Allys:2015yda}, without using a specified GUT, and considering only the bosonic structure of the strings. We continue this study by considering a given SO(10) supersymmetric (SUSY) GUT, which has already been studied in a particle physics and a cosmological framework \cite{Aulakh:2002zr,Fukuyama:2004ps,Fukuyama:2004xs,Aulakh:2004hm,Bajc:2004xe,Aulakh:2005mw,Bajc:2005qe,Aulakh:2003kg,Cacciapaglia:2013tga,Fukuyama:2012rw,Garg:2015aga}. We also consider that the SSB scheme of the GUT takes place during a $F$-term hybrid inflation scenario~\cite{Copeland:1994vg,Linde:1993cn,Dvali:1994ms,Lyth:1998xn,Kyae:2005vg,Mazumdar:2010sa}, where the inflaton dynamics is driven by the field content of the GUT itself.

Having a complete description of these linear topological defects can give interesting results. On the one hand, their macroscopic properties could considerably change in comparison with the simpler models, which would modify the current constraints using cosmic strings observations. On the other hand, it gives access to the parameters of the GUT itself, and not only to the energy scale of formation of the strings. For example, having the energy per unit length of the string as function of the different parameters of the GUT permits to implement observational constraints on their ranges.

In a first part, we introduce the SUSY SO(10) GUT studied, and its SSB scheme in parallel with the inflationary process. We then focus in Sec.~\ref{TensorFormulation} on the explicit tensor formulation of the theory. Special attention is payed to the formulation of the model as a function of the restricted representations which are singlet under the SM. Some of the calculations and results of this section are put in Appendix \ref{AppendixA}. The cosmic strings are studied in Sec.~\ref{PartCosmicStrings}, where we give an ansatz for their structure, and write the equations of motion as well as the
boundary conditions for all the fields. After writing the model with dimensionless variables, we perform a perturbative study of the string in Sec.~\ref{PartScaling&Pert}. Finally, in Sec.~\ref{PartNumericalSolution}, we present and discuss the numerical solutions of the strings, and their microscopic and macroscopic properties.

%%%%%%%%%%%%%%%%%%%%%%%%%%%%%%%%%%%%%%%%%%%%%%%%%%%%%%%%%%%%%%%%%
%%%%%%%%%%%%%%%%%%%%%%%%%%%%%%%%%%%%%%%%%%%%%%%%%%%%%%%%%%%%%%%%%
\section{SO(10) GUT, hybrid inflation, and SSB}
%%%%%%%%%%%%%%%%%%%%%%%%%%%%%%%%%%%%%%%%%%%%%%%%%%%%%%%%%%%%%%%%%
\subsection{GUT and field content}
\label{PartIntroGUT}

We focus on a well-studied SO(10) SUSY GUT, which has already been considered in a particle physics \cite{Aulakh:2002zr,Fukuyama:2004ps,Fukuyama:2004xs,Aulakh:2004hm,Bajc:2004xe,Aulakh:2005mw,Bajc:2005qe,Aulakh:2003kg,Fukuyama:2012rw} and a cosmological \cite{Cacciapaglia:2013tga,Garg:2015aga} framework. The superpotential yields \cite{Bajc:2004xe,Martin:1997ns,Slansky:1981yr}
\begin{multline}
W= \frac{m}{2}\mathbf{\Phi}^2
+ m_{\Sigma}\mathbf{\Sigma}\overline{\mathbf{\Sigma}}
+ \frac{\lambda}{3}\mathbf{\Phi}^3 \\
+ \eta\mathbf{\Phi}\mathbf{\Sigma}\overline{\mathbf{\Sigma}}
+ \kappa S (\mathbf{\Sigma}\overline{\mathbf{\Sigma}}-M^2),
\end{multline}
where $\mathbf{\Sigma}$ and $\mathbf{\overline{\Sigma}}$ are in the \textbf{126} and $\mathbf{\overline{126}}$ representations, $\mathbf{\Phi}$ is in the \textbf{210} representation, and the inflaton $S$ is a singlet of SO(10). It is the more general singlet term we can write, in addition with an $F$-term hybrid inflation coupling involving the inflaton and $\mathbf{\Sigma}$ and $\mathbf{\overline{\Sigma}}$, which are the only fields in complex conjugate representations. This last term is the simplest we can write which reproduces the inflation phenomenology \cite{Lyth:1998xn,Mazumdar:2010sa}. Additional terms in $S^2$, $S^3$ or $S \mathbf{\Phi}^2$ could be added, but they often generate mass or quartic terms for the inflaton at tree-level, and then spoil the inflation. A discussion of those terms can be found in Ref.~\cite{Cacciapaglia:2013tga}.

All the parameters introduced are complex, but we can use redefinitions of the phases of the superfields to set $m$, $m_{\Sigma}$, $\kappa$ and $M$ real, $\lambda$ and $\eta$ still being complex. The explicit component formulation will be given in Sec.~\ref{TensorFormulation}. The reader should be reminded that as we work in a SUSY framework, all the components of the scalar fields are complex.

Note that the purpose of this article is to consider in details the complete GUT structure of the strings, rather than focus on the inflation, which is treated at a basic level. We could also consider an inflation led by fields out of the GUT sector, and recover the same kind of phenomenology for the strings. The advantage here is that, inflation being implemented by the GUT fields, we have a precise relation between the cosmological evolution and the breaking scheme of the GUT, see Sec.~\ref{SSB&HI}. It is also possible to refine the model by including additional couplings to develop the inflationary phenomenology.

This superpotential is sufficient to describe the SSB of the GUT to the SM symmetry. In addition, a Higgs field in the \textbf{10} representation permits us to implement the electroweak symmetry breaking. Its characteristic scale is very different of that considered in this paper and we can omit it from now on. Another Higgs field in the \textbf{120} representation of SO(10) can be added in order to recover the complete SM fermion mass spectrum, but it can be omitted in a first approximation \cite{Aulakh:2005mw}.

%
%We chose to keep the simplest term concerning the inflation since we want to focus on the effect of a realistic GUT implementation. However, a study with a non minimal inflationary potential can also be done, see for example \cite{Kyae:2005vg} for a SUSY GUT model with a non-minimal inflation term, and \cite{Davis:1999tk} for the phenomenology of cosmic strings with a minimal Higgs content but with a many inflatons phenomenology.

From now on, we will focus on the bosonic sector of the theory, so we will not consider the superfields anymore, and focus instead on their bosonic part. Also, as we work with $F$-term hybrid inflation scenario, we assume that all the $D$-terms associated to the gauge generators take identically vanishing values, with no Fayet-Iliopoulos term \cite{Martin:1997ns}. This condition will permit us to impose some constraints on the fields thereafter.

%%%%%%%%%%%%%%%%%%%%%%%%%%%%%%%%%%%%%%%%%%%%%%%%%%%%%%%%%%%%%%%%%
\subsection{Lagrangian of the bosonic sector}

We take the signature of the metric to be $+2$. The Lagrangian of the bosonic sector gives
\begin{multline}
\label{LagFull}
\mathcal{L}=-\frac{1}{4}\text{Tr}(F_{\mu \nu}F^{\mu \nu})
-(D_{\mu}\mathbf{\Phi})^\dagger(D^{\mu}\mathbf{\Phi})\\
-(D_{\mu}\mathbf{\Sigma})^\dagger(D^{\mu}\mathbf{\Sigma})
-(D_{\mu}\mathbf{\overline{\Sigma}})^\dagger(D^{\mu}\mathbf{\overline{\Sigma}})\\
-(\nabla_{\mu}S)^\ast(\nabla^{\mu}S)
-V(\mathbf{\Phi},\mathbf{\Sigma},\mathbf{\overline{\Sigma}},S).
\end{multline}
The inflaton has no gauge covariant derivative since this is a singlet of the gauge group. We also take the following definitions
\begin{equation}
D_{\mu}X=(\nabla_{\mu}+ g A_{\mu})X,
\end{equation}
\begin{equation}
A_{\mu}=-iA_{\mu}^{a}\tau_X^a,
\end{equation}
\begin{equation}
F_{\mu\nu}=-F_{\mu\nu}^{a}\tau_X^a,
\end{equation}
\begin{equation}
F_{\mu\nu}=\nabla_\mu A_{\nu}-\nabla_\nu A_{\mu}+q [A_{\mu},A_{\nu}],
\end{equation}
where from now on $X$ is a general notation for all scalar fields, i.e. $X\in \{\mathbf{\Sigma}, \overline{\mathbf{\Sigma}}, \mathbf{\Phi}, S \}$. We note $\tau_X^a$ the action of the generators of the gauge group in the representation of $X$, the index $a$ labeling the $45$ generators of SO(10). The \textbf{210} representation being real, we can use a basis where ${(\tau_{210}^a)}^\dagger=\tau_{210}^a$. It is not the case anymore for the \textbf{126} representation, which is complex: ${(\tau_{126}^a)}^\dagger$ is not anymore in the \textbf{126} but in the $\mathbf{\overline{126}}$ representation. However, we can choose a basis where $\tau_{\overline{126}}^a=-{(\tau_{ 126}^a)}^\dagger$.
Independently of the representation, we have 
\begin{equation}
[\tau_{a},\tau_{b}]=if_{ab}{}^c\tau_{c},
\end{equation}
with $f_{ab}{}^{c}$ the structure constants of SO$(10)$.

We first focus on the $F$-term part of the potential. They are defined by
%\footnote{\label{FootnoteF-term}From the point of view of superspace, it is more suitable to use $F_X={\left({\partial W}/{\partial X}\right)}^\dagger$. Indeed, $X$ and $F_X$ have to be in the same representation in order to enter the same supermultiplet.  However, as these $F$-terms appear only in the potential with a term in $F^\dagger F $, the alternative definition of Eq. (\ref{DefFTerms}) is equivalent.}
\begin{equation}
\label{DefFTerms}
F_X=\frac{\partial W}{\partial X},
\end{equation}
with $F_X$ in the conjugate representation of $X$.
It yields
\begin{equation}
\begin{array}{l}
F_S= \kappa  (\mathbf{\Sigma}\overline{\mathbf{\Sigma}}-M^2), \\
F_{\Phi}=m\mathbf{\Phi} + \lambda(\mathbf{\Phi} ^2)_{\mathbf{\Phi}} + \eta(\mathbf{\Sigma}\overline{\mathbf{\Sigma}})_{\mathbf{\Phi}},\\
F_{\Sigma}=m_{\Sigma}\overline{\mathbf{\Sigma}}+ \eta(\mathbf{\Phi}\overline{\mathbf{\Sigma}})_{\overline{\mathbf{\Sigma}}} + \kappa S \overline{\mathbf{\Sigma}}, \\
F_{\overline{\Sigma}}=m_{\Sigma}\mathbf{\Sigma}+ \eta(\mathbf{\Phi}\mathbf{\Sigma})_{\mathbf{\Sigma}} + \kappa S \mathbf{\Sigma},
\end{array}
\end{equation}
where we denote with $(X Y)_Z$ the term in the representation of $Z$ we can build from the product of the fields $X$ and $Y$. See Sec.~\ref{AppendixDerivatives} for a more detailed explanation about how to obtain these results.

We finally obtain for the potential
\begin{equation}
V=\sum_X F_X^\dagger F_X\equiv\sum_X V_X,
\end{equation}
which is a sum of positive terms. 
It gives
\begin{equation}
\label{PotentialS}
V_S= \kappa^2 (\mathbf{\Sigma}\overline{\mathbf{\Sigma}}-M^2)^2,
\end{equation}
\begin{align}
V_\Phi=&m^2\mathbf{\Phi} \mathbf{\Phi}^\dagger 
+ |\eta|^2 (\mathbf{\Sigma}\overline{\mathbf{\Sigma}})_{\mathbf{\Phi}} (\mathbf{\Sigma}\overline{\mathbf{\Sigma}})_{\mathbf{\Phi}}^\dagger \nonumber \\
& + |\lambda|^2 (\mathbf{\Phi} ^2)_{\mathbf{\Phi}} (\mathbf{\Phi} ^2)_{\mathbf{\Phi}} ^\dagger 
+ \left[ \lambda \eta^* (\mathbf{\Phi} ^2)_{\mathbf{\Phi}}(\mathbf{\Sigma}\overline{\mathbf{\Sigma}})_{\mathbf{\Phi}}^\dagger + 
\right.  \nonumber \\
&  \label{PotentialPhi} \left. + m \lambda^* \mathbf{\Phi}(\mathbf{\Phi} ^2)_{\mathbf{\Phi}}^\dagger  
+ m \eta^* \mathbf{\Phi}(\mathbf{\Sigma}\overline{\mathbf{\Sigma}})_{\mathbf{\Phi}}^\dagger \right] + \text{h.c.},
\end{align}

\begin{align}
\label{PotentialSigma}
V_{\overline{\Sigma}}=& m_{\Sigma}^2 \mathbf{\Sigma}\mathbf{\Sigma}^\dagger   + |\eta|^2 (\mathbf{\Phi}\mathbf{\Sigma})_{\mathbf{\Sigma}} (\mathbf{\Phi}\mathbf{\Sigma})_{\mathbf{\Sigma}}^\dagger
+ \kappa^2 SS^*\mathbf{\Sigma}\mathbf{\Sigma}^\dagger \nonumber \\
&+\left[ \eta \kappa S^{*} (\mathbf{\Phi}\mathbf{\Sigma})_{\mathbf{\Sigma}} \mathbf{\Sigma}^\dagger 
+ m_{\Sigma} \kappa S^*  \mathbf{\Sigma}\mathbf{\Sigma}^\dagger \right. \nonumber \\ 
& \left. +  m_{\Sigma}  \eta^*\mathbf{\Sigma}(\mathbf{\Phi}\mathbf{\Sigma})_{\mathbf{\Sigma}}^{\dagger}\right] +\text{h.c.},
\end{align}
and
\begin{equation}
\label{PotentialSigmaBar}
V_{\Sigma}=V_{\overline{\Sigma}} (\mathbf{\Sigma} \longleftrightarrow \mathbf{\overline{\Sigma}} ).
\end{equation}
Note that we did not include in what is called $V$ the $D$-term contribution to the potential. Indeed, this term having an identically vanishing value, it does not contribute to the dynamics of the fields. However, this condition will be imposed by the following, and gives some constraints on the fields.

%%%%%%%%%%%%%%%%%%%%%%%%%%%%%%%%%%%%%%%%%%%%%%%%%%%%%%%%%%%%%%%%%
\subsection{SSB scheme, hybrid inflation and topological defects}
\label{SSB&HI}

We now turn to the cosmological evolution of the GUT, following Refs.~\cite{Cacciapaglia:2013tga,Allys:2015yda}, and also Ref.~\cite{Jeannerot:1995yn} for a model very close to the one we study. The SSB schemes take the following form
\begin{equation}
\text{SO}(10)\overset{\langle\Phi\rangle}{\relbar\joinrel\relbar\joinrel\longrightarrow}
G' \overset{\langle\Sigma\rangle}{\relbar\joinrel\relbar\joinrel\longrightarrow}
G_{\text{SM}} \times \mathbf{Z}_2,
\end{equation}
where $G_{\text{SM}}=3_C 2_L 1_Y$. We use from now on short notations for the gauge groups, $3_C 2_L 1_Y$ meaning SU(3)$_C \times$SU(2)$_L \times$U(1)$_Y$, and so on. The $\mathbf{Z}_2$ factor appears in addition to the SM gauge group in order to suppress proton decay \cite{Martin:1992mq}. We assume that the first SSB step happens at $E_{\text{GUT}}\sim 10^{16}$GeV, and the second one slightly below, at $E\sim \left(10^{15}-10^{16}\right)$GeV.

At the onset of inflation, we can assume a very large value for the inflaton $S$ in comparison with all the other fields, as it is the case in chaotic inflation. To minimize the higher order terms containing it in Eqs.~(\ref{PotentialSigma}) and~(\ref{PotentialSigmaBar}), so that $V_{\bar{\Sigma}}\sim |\kappa S \mathbf{\Sigma}|^2$ and $V_\Sigma \sim |\kappa S \overline{\mathbf{\Sigma}}|^2$, the field $\mathbf{\Sigma}$ and $\mathbf{\overline{\Sigma}}$ must take a vanishing value. We thus have $V_{\Sigma}=0$, $V_{\overline{\Sigma}}=0$ and $V_S=\kappa^2 M^4$ at this step. In addition, we assume that the VEV taken by $\mathbf{\Phi}$ minimizes $V_\Phi$. Different VEVs for $\mathbf{\Phi}$ have this property, each one being associated with a different gauge group at this step $G'$.

After this first SSB, a false vacuum hybrid inflation begins. Indeed, the potential verifies $V=V_0 + \text{quant}.\text{corr}.$, with $V_0=\kappa^2 M^4$ and the inflaton slowly rolls along this flat direction at tree level until it reaches its critical value, thus ending inflation (see e.g. Ref.~\cite{Cacciapaglia:2013tga} for a detailed explanation about this phase). At this point, a new phase transition takes place, causing a SSB down to the SM gauge group. We assume the set of VEVs at this last step to be a global minimum for the potential, i.e. implying $V=0$. As previously, this condition is fulfilled by several set of VEVs, each one defining an unbroken gauge symmetry.

A careful study of the possible SSB cascades has been done in \cite{Cacciapaglia:2013tga}, including stability of the inflationary valley. Only two of them are valid and permit us to recover the SM at low energy. Their respective intermediate symmetry group is $G'=3_C 2_L 2_R 1_{B-L}$ and $G'=3_C 2_L 1_R 1_{B-L}$. For both these intermediate symmetry groups, the topological defects produced are monopoles in the first SSB step, and cosmic strings in the second \cite{Jeannerot:2003qv}. The monopoles are washed out by the inflation, which takes place after their formation. In this paper, we study the cosmic strings which form in the SSB from $G'$ to $G_{\text{SM}}\times \mathbf{Z}_2$, at the end of inflation.

%%%%%%%%%%%%%%%%%%%%%%%%%%%%%%%%%%%%%%%%%%%%%%%%%%%%%%%%%%%%%%%%%
\subsection{Use of fields singlet under the SM}
\label{PartSMSinglet}

The VEVs which have a non vanishing value at the last stage of SSB define the SM symmetry. It implies that they must be uncharged under $G_{\text{SM}}$, since they would otherwise break this symmetry group at this stage. These non vanishing VEVs thus are singlet under the SM. We will also assume there is no symmetry restoration, i.e. all fields acquiring a non-zero VEV at a given step keep it non vanishing at later stages. It permits us to impose that all the non vanishing VEVs of the SSB cascade are also singlet under the SM.

We can consider all the restricted representations of the field content of the GUT we study, and look for sub-representations which are singlet under the SM. These are very few \cite{Bajc:2004xe,Slansky:1981yr}, and are listed below, giving their representations under the Pati-Salam group ($2_L 2_R 4_C$)
\begin{equation}
\label{SMSinglet}
\begin{matrix}
   \mathbf{\Phi}_p=\mathbf{\Phi}(1,1,1), & \boldsymbol\sigma=\mathbf{\Sigma}(1,3,\overline{10}),  \\
   \mathbf{\Phi}_a=\mathbf{\Phi}(1,1,15), & \overline{\boldsymbol\sigma}=\overline{\mathbf{\Sigma}}(1,3,10), \\
   \mathbf{\Phi}_b=\mathbf{\Phi}(1,3,15), & S=S(1,1,1).
\end{matrix}
\end{equation}
Considering the \textbf{3} representation of $2_R$, it is sufficient to take the neutral component of $1_R \subset 2_R$, which is what we will do from now on. For the representations non singlet under $4_C$, their branching rules on $3_C 1_{B-L}$ are
\begin{equation}
\label{4RSingletDecomposition}
\begin{array}{l}
\mathbf{1}=\mathbf{1}(0),\\
\mathbf{10}=\mathbf{1}(2)+\mathbf{3}(2/3)+\mathbf{6}(-2/3),\\
\mathbf{15}=\mathbf{1}(0)+\mathbf{3}(-4/3)+\mathbf{\overline{3}}(4/3)+\mathbf{8}(0),
\end{array}
\end{equation}
where we denote with $\mathbf{n}(q)$ the representation of $3_C$ of dimensions $n$ which has a charge $q$ under $1_{B-L}$.
For these representations, we use their sub-representation singlet of $3_C$, which is unique. The explicit tensor formulation of these restricted representations will be given in Sec.~\ref{PartSingletDecomposition}.

It permits us to describe in a short way the non vanishing VEVs appearing after the first SSB scheme and defining $G'$, for the two relevant schemes \cite{Bajc:2004xe,Cacciapaglia:2013tga}. For the first one, only $\langle \mathbf{\Phi}_a \rangle$ takes a non vanishing value, and the residual symmetry group is $G'_1=3_C 2_L 2_R 1_{B-L}$. For the second one, the three restricted representations singlet under the SM contained in $\mathbf{\Phi}$ take a non vanishing expectation value, and the residual symmetry group is $G'_2=3_C 2_L 1_R 1_{B-L}$.
%Note that the $1_{B-L}$ symmetry is always unbroken at this step; indeed, all the singlets under the SM contained in $\mathbf{\Phi}$ are uncharged under this symmetry.

Finally, and as discussed in Ref.~\cite{Allys:2015yda}, we can restrict the study of the microscopic structure of the string to a configuration where all the fields which are not singlet under the SM take an identically vanishing value. Indeed, the potential is at least quadratic in these fields, since it would otherwise be charged under the SM. So the solutions where all these fields take a vanishing value is a solution of their equations of motion. On the other hand, these fields must have a zero value at infinity since the vacuum is uncharged under the SM. This shows that the solution discussed previously is also compatible with the boundary conditions at infinity. We will assume this particular ansatz from now on. For a more detailed discussion about this assumption, see Ref.~\cite{Allys:2015yda}.

%%%%%%%%%%%%%%%%%%%%%%%%%%%%%%%%%%%%%%%%%%%%%%%%%%%%%%%%%%%%%%%%%
%%%%%%%%%%%%%%%%%%%%%%%%%%%%%%%%%%%%%%%%%%%%%%%%%%%%%%%%%%%%%%%%%
\section{Explicit tensor formulation}
\label{TensorFormulation}
%%%%%%%%%%%%%%%%%%%%%%%%%%%%%%%%%%%%%%%%%%%%%%%%%%%%%%%%%%%%%%%%%
\subsection{Introduction and fields content}

In order to have a complete study of the model, including numerical solution, we need to write it in a component (here tensor) formulation, which is the purpose of this whole section. As explained in Sec \ref{PartSMSinglet}, we will also restrict the study to only the fields which are singlet under the SM, which permits us to describe the problem in terms of only a few complex functions. In order to be as concise and clear as possible, a part of the calculations and results are given as an appendix, in Sec.~\ref{AppendixA}.

The tensor formulation of the field content yields \cite{Bajc:2004xe,Aulakh:2000sn} : 
\begin{itemize}
\item $\mathbf{\Sigma}$ (\textbf{126}) is a fifth rank anti-symmetric tensor $\Sigma_{ijklm}$, self dual (in the sense of Hodge duality) :
\begin{equation}
\label{self-dual}
\Sigma_{ijklm}=\frac{i}{5!}\epsilon_{ijklmabcde}\Sigma_{abcde},
\end{equation}
\item $\overline{\mathbf{\Sigma}}$ ($\mathbf{\overline{126}}$) is a fifth rank anti-symmetric tensor $\overline{\Sigma}_{ijklm}$, anti-self-dual :
\begin{equation}
\overline{\Sigma}_{ijklm}=-\frac{i}{5!}\epsilon_{ijklmabcde}\overline{\Sigma}_{abcde},
\end{equation}
\item $\mathbf{\Phi}$ (\textbf{210}) is a fourth rank anti-symmetric tensor $\Phi_{ijkl}$,
\item $S$ is a singlet
\end{itemize}
We remind the reader that in the tensor formulation of SO(10), all the indices go between $1$ and $10$.
 
%%%%%%%%%%%%%%%%%%%%%%%%%%%%%%%%%%%%%%%%%%%%%%%%%%%%%%%%%%%%%%%%%
\subsection{Superpotential, $F$-terms, and potential}
\label{TensorW&F}

The superpotential, defined in Sec.~\ref{PartIntroGUT}, yields
\begin{multline}
W= \frac{1}{2}m \Phi_{ijkl} \Phi_{ijkl}
+ m_{\Sigma} \Sigma_{ijklm} \overline{\Sigma}_{ijklm}\\
+ \frac{1}{3}\lambda\Phi_{ijkl} \Phi_{klmn}\Phi_{mnij} 
+ \eta \Phi_{ijkl}\Sigma_{ijmno} \overline{\Sigma}_{klmno}\\
+ \kappa S (\Sigma_{ijklm} \overline{\Sigma}_{ijklm}-M^2)
\end{multline}

Now, to compute the $F$-terms, we have to take the derivatives with respect to the different tensor components of the fields. However, we have to take into account the fact that they are not independent (due to the symmetry and duality properties of the tensors). The way to proceed is given in Sec.~\ref{AppendixDerivativeComputation}, and the derivatives of the different terms and the associated notations are written in Sec.~\ref{AppendixDerivatives}. Explicitly, we find
\begin{equation}
\displaystyle{F_S= \kappa  (\Sigma_{ijklm} \overline{\Sigma}_{ijklm}-M^2)},
\end{equation}
\begin{multline}
{\left(F_{\Phi}\right)}_{ijkl}= m \Phi_{ijkl} + \lambda \Phi_{[ij|ab}\Phi_{ab|kl]}\\
 + \eta \Sigma_{[ij|abc}\overline{\Sigma}_{|kl]abc},
\end{multline}
\begin{multline}
{\left(F_{\Sigma}\right)}_{ijklm}=m_{\Sigma} \overline{\Sigma}_{ijklm} + \frac{\eta }{2}\bigg(\Phi_{[ij|\alpha\beta}\overline{\Sigma}_{\alpha\beta |klm]} \\
-\frac{i}{5!}\epsilon_{ijklmabcde}\Phi_{ab\alpha\beta}\overline{\Sigma}_{\alpha\beta cde}\bigg)  + \kappa S  \overline{\Sigma}_{ijklm},
\end{multline}
and
\begin{multline}
{\left(F_{\overline{\Sigma}}\right)}_{ijklm}=m_{\Sigma} \Sigma_{ijklm} +  \frac{\eta}{2}\bigg(\Phi_{[ij|\alpha\beta}\Sigma_{\alpha\beta |klm]}\\
+\frac{i}{5!}\epsilon_{ijklmabcde}\Phi_{ab\alpha\beta}\Sigma_{\alpha\beta cde}\bigg) + \kappa S  \Sigma_{ijklm}.
\end{multline}
Note that in these $F$-terms, we cannot obtain $F_{\overline{\Sigma}}$ from $F_{\Sigma}$ by only changing $\Sigma_{abcde}$ in $\overline{\Sigma}_{abcde}$.

We do not write the full tensorial expression of the potential at this step, which is obtained by injecting the results of Sec.~\ref{AppendixDerivatives} in Eq.~(\ref{PotentialS}) to (\ref{PotentialSigmaBar}).
 
%%%%%%%%%%%%%%%%%%%%%%%%%%%%%%%%%%%%%%%%%%%%%%%%%%%%%%%%%%%%%%%%%
\subsection{Singlet decomposition and $D$-terms}
\label{PartSingletDecomposition}

Let us consider now the restricted representations which are singlet under the SM. In addition, they are uncharged under any continuous non abelian symmetry which commutes with the SM symmetry. It implies that we can describe their dynamics by a single complex function. It yields, e.g. 
\begin{equation}
\mathbf{\Phi}_{a}  = a(x^\mu) {\langle \mathbf{\Phi}_{a}\rangle}_0,
\end{equation}
where $a(x^\mu)$ is a complex function of the space-time, and ${\langle \mathbf{\Phi}_{a}\rangle}_0$ is a constant vector in the representation space. Following the conventions of \cite{Allys:2015yda}, we choose to work with normalized constant vectors, i.e. with ${\langle \mathbf{\Phi}_{a}\rangle}_0 {\langle \mathbf{\Phi}_{a}\rangle}_0^\dagger=1$. 

We can now write these singlets under the SM in a tensor formulation \cite{Bajc:2004xe}, following the notations introduced in Eq.~(\ref{SMSinglet}),
\begin{equation}
\label{SingletDefinition}
\left\{
\begin{array}{l}
\displaystyle{ \frac{p}{\sqrt{4!}}= \Phi_{1234} },\\
\displaystyle{\frac{a}{\sqrt{4!3}}= \Phi_{5678} = \Phi_{5690} 
= \Phi_{7890} },\\
\displaystyle{\frac{b}{\sqrt{4!6}}= \Phi_{1256} = \Phi_{1278} 
= \Phi_{1290} }\\
\displaystyle{~~~~~~~ = \Phi_{3456} 
= \Phi_{3478} = \Phi_{3490} },\\
\displaystyle{\frac{1}{\sqrt{5!2^5}}(i)^{(-a-b+c+d+e)}\sigma}\\
~~~~ ~~~~ ~~~~ ~~~~ ~~~~ =\Sigma_{a+1,b+3,c+5,d+7,e+9}, \\
\displaystyle{\frac{1}{\sqrt{5!2^5}}(-i)^{(-a-b+c+d+e)}\overline{\sigma}}\\
~~~~ ~~~~ ~~~~ ~~~~ ~~~~ = \overline{\Sigma}_{a+1,b+3,c+5,d+7,e+9},
\end{array}
\right.
\end{equation}
where the complex functions are $p$, $a$, $b$, $\sigma$ and $\overline{\sigma}$. In the last two equations, the indices $a$, $b$, $c$, $d$ and $e$ are either $0$ or $1$.

The $D$-term condition permits us to impose additional constraints  on the fields. The general expression is
\begin{equation}
\label{D-term}
D^a=-g\sum_X (X^\dagger \tau^a_X X),
\end{equation}
for the $D$-term associated to the generator $\tau^a$ (in the case of a vanishing Fayet-Iliopoulos term) \cite{Martin:1997ns}. The associated potential in the Lagrangian is
\begin{equation}
V_{D} = \frac{1}{2} \sum_a D^a D^a
\end{equation}
As we work in the framework of an $F$-term theory, all the $D$-terms identically vanish. The $D$-term condition associated to the generator of $1_{B-L}$ simplifies a lot since only the SM singlets associated to $\mathbf{\Sigma}$ and $\mathbf{\overline{\Sigma}}$ are charged under this group [see Eqs.~\ref{4RSingletDecomposition}], yielding \cite{Slansky:1981yr,Harada:2003sb}
% This result can also been seen explicitly in Eq.~(\ref{SingletDefinition}), since the generator associated with $1_{B-L}$ implies simultaneous rotations in the $1-2$, $3-4$, $5-6$, $7-8$ and $9-0$ planes \cite{Ma:1992ky}. 
\begin{equation}
D_{1_{B-L}} = \sqrt{\frac{3}{8}}\left( 2{ \boldsymbol\sigma }^\dagger \boldsymbol\sigma  -2 { \overline{\boldsymbol\sigma} }^\dagger \overline{\boldsymbol\sigma}\right)  = 0.
\end{equation}
In addition, as the phase of the inflaton $S$ have been rephased in order to make $M$ real, it ensures that the global minimum of the potential is reached when $\boldsymbol\sigma \overline{\boldsymbol\sigma}=M^2 \in \mathbb{R}$. Both these results impose that $\overline{\boldsymbol\sigma}=\boldsymbol\sigma^\dagger$, which finally gives $\overline{\sigma}=\sigma^*$.

To summarize, we can write the fields in the VEV directions defined in Eq.~(\ref{SingletDefinition}) 
\begin{equation}
\left\{
\begin{array}{l}
\displaystyle{ \langle \boldsymbol\Sigma \rangle ={\langle \overline{\boldsymbol\Sigma} \rangle }^\dagger =  \sigma {\langle \boldsymbol\sigma \rangle }_0},\\
\displaystyle{\langle \mathbf{\Phi} \rangle = a {\langle \mathbf{\Phi}_a \rangle}_0 + b {\langle \mathbf{\Phi}_b \rangle}_0 + p {\langle \mathbf{\Phi}_p \rangle}_0 },
\end{array}
\right.
\end{equation}
with the normalization conditions giving
\begin{equation}
\label{Orthonormalization}
\begin{array}{l}
{\langle \mathbf{\Phi}_p\rangle}_0 {\langle \mathbf{\Phi}_p \rangle}_0^\dagger =1,\\
{\langle \mathbf{\Phi}_a \rangle}_0 {\langle \mathbf{\Phi}_a \rangle}_0^\dagger =1,\\
{\langle \mathbf{\Phi}_b \rangle}_0 {\langle \mathbf{\Phi}_b \rangle}_0^\dagger =1,\\
\displaystyle{ {\langle \boldsymbol\sigma \rangle }_0 {\langle \boldsymbol\sigma \rangle }_0^\dagger = {\langle \boldsymbol\sigma \rangle }_0{\langle \overline{\boldsymbol\sigma} \rangle }_0 =1.}
\end{array}
\end{equation}
The other scalar products vanish, since we cannot construct scalar terms with two fields which are not in conjugate representations.

Finally, the different contractions between the fields written in terms of the few complex functions introduced previously can be found in Appendix~\ref{AppendixSelectionRules}.

%%%%%%%%%%%%%%%%%%%%%%%%%%%%%%%%%%%%%%%%%%%%%%%%%%%%%%%%%%%%%%%%%
\subsection{Superpotential and potential in singlet form}
\label{PartW&VSinglet}

We can now write the superpotential and the $F$-term scalar potential terms in term of the few complex functions introduced in the previous section. The superpotential gives 
\begin{multline}
W=\frac{m}{2}\left(p^2+a^2+b^2 \right)+m_{\Sigma}\sigma\sigma^*+\frac{\lambda}{3}\left(\frac{a^3}{9\sqrt{2}}
\right.\\ \left.
+\frac{ab^2}{3\sqrt{2}} +\frac{pb^2}{2\sqrt{6}}\right) +\eta \sigma \sigma^*\left(\frac{p}{10\sqrt{6}}+\frac{a}{10\sqrt{2}}-\frac{b}{10} \right)\\
+\kappa S \left( \sigma \sigma^* -M^2\right).
\end{multline}
In a similar way, the potential can be written by using the expressions given in Sec.~\ref{AppendixSelectionRules} in the Eq.~(\ref{PotentialS}) to (\ref{PotentialSigmaBar}). However, it is possible to write it in a more convenient form.

For this purpose, we can introduce the $F$-terms associated with the restricted representations singlet under the SM. Indeed, the only non vanishing terms in the $F$-term associated to $\mathbf{\Phi}$ are
\begin{equation}
\label{EqFp}
\frac{F_p}{2 \sqrt{6}}={\left(F_{\Phi}\right)}_{1,2,3,4}=\frac{m p}{2 \sqrt{6}} + \frac{\lambda b^2}{72} +\frac{ \eta\sigma\sigma^*}{120},
\end{equation}
\begin{multline}
\label{EqFa}
\frac{F_a}{6 \sqrt{2}}={\left(F_{\Phi}\right)}_{5,6,7,8}={\left(F_{\Phi}\right)}_{5,6,9,10}={\left(F_{\Phi}\right)}_{7,8,9,10}\\
=\frac{m a}{6\sqrt{2}}+\frac{\lambda}{3}\left( \frac{a^2}{36} + \frac{b^2}{36} \right) + \frac{\eta \sigma \sigma^*}{120},
\end{multline}
and
\begin{multline}
\label{EqFb}
\frac{F_b}{12}={\left(F_{\Phi}\right)}_{1,2,5,6}={\left(F_{\Phi}\right)}_{1,2,7,8}={\left(F_{\Phi}\right)}_{1,2,9,10}\\
={\left(F_{\Phi}\right)}_{3,4,5,6}={\left(F_{\Phi}\right)}_{3,4,7,8}={\left(F_{\Phi}\right)}_{3,4,9,10}\\
= \frac{m b}{12}+\frac{\lambda}{3}\left( \frac{ab}{18\sqrt{2}} + \frac{bp}{12\sqrt{6}} \right) - \frac{\eta \sigma \sigma^*}{120}.
\end{multline}
$F_{\Phi}$ being in the same representation as $\mathbf{\Phi}$, they can be identified as the terms in the representations of $\mathbf{\Phi}_p$, $\mathbf{\Phi}_a$ and $\mathbf{\Phi}_b$ appearing in its branching rules. We also introduce, without specifying anymore all the sets of indices obtained by considering the antisymmetric and self-dual configurations [see Eq.~(\ref{SingletDefinition})],
%\footnote{Note that $F_{\overline{\Sigma}}$ is in the representation of $\mathbf{\Sigma}$, see the footnote \ref{FootnoteF-term} associated to the definition of the $F$-terms before Eq. (\ref{DefFTerms}).}
\begin{multline}
\frac{F_\sigma}{16\sqrt{15}}={\left(F_{\overline{\Sigma}}\right)}_{1,3,5,7,9}= \frac{m_\Sigma \sigma}{16\sqrt{15}} \\ +\frac{\eta  \sigma}{960\sqrt{5}}\left( \sqrt{6}a 
-2\sqrt{3}b+\sqrt{2}p \right)+  \frac{\kappa S\sigma}{16\sqrt{15}}.
\end{multline}
Finally, we have
\begin{equation}
F_S = \kappa (\sigma\sigma^* - M^2 ).
\end{equation}

These definitions permit us to write the $F$-term scalar potential in a simpler form,
\begin{equation}
\label{EqPotWithFterms}
\begin{split}
V & = V_\Phi + V_\Sigma + V_{\overline{\Sigma}} + V_S \\
&= {F_{\Phi}}{F_{\Phi}}^\dagger + {F_{\Sigma}}{F_{\Sigma}}^\dagger + {F_{\overline{\Sigma}}}{F_{\overline{\Sigma}}}^\dagger + F_S F_S^*\\
&= \left| F_p\right|^2+\left| F_a\right|^2+\left| F_b\right|^2 + 2 \left| F_\sigma \right|^2 + |F_S|^2.
\end{split}
\end{equation}
This formulation of the potential is useful since it shows explicitly the sum of positive terms. So, when doing only a static study of the problem as done in Sec.~\ref{SSB&HI}, i.e. when not comparing the different terms, it is possible to work with these few simplified $F$-terms only, as often done in the literature. Note that the $F$-terms can indeed be recovered from the usual definition 
\begin{equation}
F_a = \frac{\partial W}{\partial a},
\end{equation}
and so on. However, the simple form of the potential given in Eq.(\ref{EqPotWithFterms}) is permitted only because we chose normalized fields. 

Before going on, let us mention that other papers use different conventions when defining the superpotential and the kinetic part of the Lagrangian, as well as the singlet configurations in tensor formulation (which can be not normalized). The normalization choice we use is useful to properly recover the abelian Higgs model in a given limit discussed in Sec.~\ref{PartOdGToyModel}. We give in Sec.~\ref{AppendixAlternativeFormulation} the link between the expressions of the present paper, and the expressions found in \cite{Aulakh:2003kg,Bajc:2004xe,Cacciapaglia:2013tga}. All the results obtained are indeed compatible with the previous works on the subject.

%%%%%%%%%%%%%%%%%%%%%%%%%%%%%%%%%%%%%%%%%%%%%%%%%%%%%%%%%%%%%%%%%
%%%%%%%%%%%%%%%%%%%%%%%%%%%%%%%%%%%%%%%%%%%%%%%%%%%%%%%%%%%%%%%%%
\section{Abelian cosmic strings}
\label{PartCosmicStrings}
%%%%%%%%%%%%%%%%%%%%%%%%%%%%%%%%%%%%%%%%%%%%%%%%%%%%%%%%%%%%%%%%%
\subsection{Introduction, cosmic strings studied}
\label{PartIntroCosmicStrings}

We now turn to the study of the cosmic strings which form at the second step of SSB which ends hybrid inflation, at $E\sim \left(10^{15}-10^{16}\right)$GeV (see Sec.~\ref{SSB&HI}). As we saw from a cosmological study, the two possible SSBs at this step are from  $G'_1=3_C 2_L 2_R 1_{B-L}$ or $G'_2=3_C 2_L 1_R 1_{B-L}$ to $G_{\text{SM}}\times\mathbf{Z}_2$. In both cases, only cosmic strings form at this stage.
As detailed in \cite{Allys:2015yda}, these strings cannot connect with monopoles. From now on, we use a set of cylindrical coordinates $(r, \theta, z, t)$ based on the location of the string, and taken to be locally aligned along the $z$-axis ar $r=0$. We also focus on strings with fields which are functions only of $r$ and $\theta$, i.e. Nambu-Goto strings. 

In the case of $G'_2 \rightarrow G_{\text{SM}}\times\mathbf{Z}_2$, only abelian strings associated with the generator of $1_{B-L}$ can form. But in the other case, $G'_1 \rightarrow G_{\text{SM}}\times\mathbf{Z}_2$, other non abelian-strings could also form, see e.g. Refs.~\cite{Aryal:1987sn,Ma:1992ky,Davis:1996sp}. We will focus in both cases on the abelian strings which could form, associated with the generator of $1_{B-L}$. As explained in Sec.~\ref{PartSingletDecomposition}, only the non-zero VEVs associated to $\mathbf{\Sigma}$ and $\mathbf{\overline{\Sigma}}$ are charged under this abelian group. As these fields are also the fields which are in conjugate representations and coupled with the inflaton in the superpotential, these strings are called single field strings following \cite{Allys:2015yda}.

%As this string is associated with the generator of an U(1) group, we could call it an U(1)-string. However, in order to distinguish it from the standard abelian Higgs model, to which the U(1)-strings often reduces, we prefer not to use this name. Indeed, the aim of this paper is to show that the realistic structure is actually different. 

%%%%%%%%%%%%%%%%%%%%%%%%%%%%%%%%%%%%%%%%%%%%%%%%%%%%%%%%%%%%%%%%%
\subsection{Ansatz and equation of motion}
\label{Ansatz}

In order to have unified notation with \cite{Allys:2015yda}, we call U$(1)_{\text{str}}=1_{\text{str}}$ the abelian symmetry related with the cosmic string ($1_{\scriptscriptstyle{(B-L)}}$ here), and $\tau_{\text{str}}$ the associated generator. As a first part of the ansatz, we assume that all the fields which are not singlet under the SM take an identically vanishing value, since it verifies their equations of motion, as discussed in Sec.~\ref{PartSMSinglet}. Then, we also assume that the only gauge field which does not vanish is the one associated with this generator $\tau_{\text{str}}$, which forms the string \cite{Aryal:1987sn,Ma:1992ky,Peter:1992dw,Hindmarsh:1994re,Allys:2015yda}. In order to simplify the notation, we normalize the charges associated to $1_{\text{str}}=1_{B-L}$ to have $q_\Sigma=1$ and $q_{\overline{\Sigma}}=-1$. Thus, the kinetic term, yields [using Eq.~(\ref{Orthonormalization})]
%\begin{multline}
%K= - \left|\left(\nabla_\mu-ig A_\mu^{\scriptscriptstyle{\left(B-L\right)}}\right)\sigma\right|^{2}
% - \left|\left(\nabla_\mu p\right)\right|^{2} \\
% - \left|\left(\nabla_\mu a\right)\right|^{2} - \left|\left(\nabla_\mu b\right)\right|^{2} - \left|\left(\nabla_\mu S\right)\right|^{2} \\
%  - \frac{\text{Tr}( { \tau^{ \scriptscriptstyle{\left(B-L\right)} } }^2)}{4}F_{\mu \nu}^{\scriptscriptstyle{\left(B-L\right)}} F^{\mu \nu \, \scriptscriptstyle{\left(B-L\right)} }.
%\end{multline}
\begin{multline}
K= - 2\left|\left(\nabla_\mu-ig A_\mu^{\text{str}}\right)\sigma\right|^{2}
 - \left|\left(\nabla_\mu p\right)\right|^{2}  - \left|\left(\nabla_\mu a\right)\right|^{2} \\
 - \left|\left(\nabla_\mu b\right)\right|^{2} - \left|\left(\nabla_\mu S\right)\right|^{2} 
  - \frac{\text{Tr} \left({\tau_{\text{str}}}^2 \right) }{4}F_{\mu \nu}^{\text{str}} F^{\mu \nu \,\text{str}}.
\end{multline}
The potential written in terms of the complex functions describing the dynamic of the singlets of the SM can be found in Sec.~\ref{PartW&VSinglet}.
%One can then verify that the Lagrangian obtained with both these contributions is invariant under the local $1_{\text{str}}$ transformation induced by 
%\begin{equation}
%\begin{array}{l}
%\sigma \longrightarrow \sigma \text{e}^{i\Omega},\\
%\sigma^* \longrightarrow \sigma^* \text{e}^{-i\Omega},\\
%A_\mu^{\text{str}} \longrightarrow A_\mu^{\text{str}} +\frac{1}{g}\partial_\mu \Omega.
%\end{array}
%\end{equation} 

The complete form of the ansatz is \cite{Aryal:1987sn,Ma:1992ky,Peter:1992dw,Hindmarsh:1994re,Allys:2015yda}
\begin{equation}
\left\{
\begin{array}{l}
p=p(r),\\
a=a(r),\\
b=b(r),\\
\sigma=f(r)e^{i n \theta},\\
\displaystyle{A_\mu=A_{\theta}^{\text{str}}(r)\tau^{\text{str}}\delta_\mu^\theta },\\
S=S(r),
\end{array}
\right.
\end{equation}
where the integer $n$ is the winding number. In this ansatz, $f$ and $Q$ are real fields, while $a$, $b$, $p$ and $S$ are complex. This ansatz gives the following equations of motion
\begin{equation}
\begin{array}{l}
\displaystyle{2 \left( f''+\frac{f'}{r}\right) =\frac{fQ^2}{r^2}+\frac{1}{2}\frac{\partial V}{\partial f}},\\ 
\displaystyle{p''+\frac{p'}{r}=\frac{\partial V}{\partial p^*}},\\[6pt] 
\displaystyle{a''+\frac{a'}{r}=\frac{\partial V}{\partial a^*}},\\ [6pt] 
\displaystyle{b''+\frac{b'}{r}=\frac{\partial V}{\partial b^*}},\\ [6pt] 
 \displaystyle{S''+\frac{S'}{r}=\frac{\partial V}{\partial S^*}},\\
\displaystyle{\text{Tr} \left({\tau_{\text{str}}}^2 \right) \left(Q''-\frac{Q'}{r}\right)=2 g^2 f^2Q},
\end{array}
\end{equation}
where a prime means a derivative with respect to the radial
coordinate $'\equiv \text{d}/\text{d}r$, and where we introduced the field
\begin{equation}
\label{defQ}
Q(r)=n-gA_\theta^{\text{str}}(r),
\end{equation}
which is a real field function of $r$ only. Note that this whole ansatz is the minimal structure one, which is developed in \cite{Allys:2015yda}.

Finally, we can reduce the model to the following effective Lagrangian
\begin{multline}
\label{LagEffDim}
\mathcal{L}_{eff}= - 2{f'}^2 - \frac{\text{Tr} \left({\tau_{\text{str}}}^2 \right) }{g^2}\frac{Q'^2}{2 r^2}- s'^\ast s - p'^\ast p' \\
  -  a'^\ast a' - b'^\ast b' -\frac{f^2Q^2}{r^2}-V(\sigma,a,b,p,s).
\end{multline}
 
%%%%%%%%%%%%%%%%%%%%%%%%%%%%%%%%%%%%%%%%%%%%%%%%%%%%%%%%%%%%%%%%%
\subsection{Boundary conditions}
\label{PartBoundaryConditions}

Let consider first the boundary conditions at infinity. For the scalar fields, they take the non vanishing VEVs discussed in Sec.~\ref{SSB&HI}, which are a global minimum of the potential and define the SM gauge symmetry. Their values will be given in the Sec.~\ref{PartScaling} in a dimensionless form. For the gauge field, we have
\begin{equation}
\lim_{r \to \infty} A_{\theta}^{\text{str}}(r)=\frac{n}{g},
\end{equation}
in order to properly cancel $D_\mu \sigma$ at infinity, which gives in term of the field $Q$
\begin{equation}
\lim_{r \to \infty} Q(r)=0.
\end{equation}

Concerning the values of the fields at the center of the strings, topological arguments and symmetry considerations give  \cite{Hindmarsh:1994re,Kibble:1976sj} 
\begin{equation}
\begin{array}{l}
f(0)=0,\\
Q(0)=n.
\end{array}
\end{equation}
For the other fields, the cylindrical symmetry around the string gives, assuming that they have a non vanishing value at the center of the string,
\begin{equation}
\frac{\text{d} x}{\text{d} r}(0)=0,
\end{equation}
for $x=p,a,b$, and $S$.

At this point, we have the equations of motion and the boundary conditions for the whole field content of the model. So it is completely defined in a mathematical point of view. 

%%%%%%%%%%%%%%%%%%%%%%%%%%%%%%%%%%%%%%%%%%%%%%%%%%%%%%%%%%%%%%%%%
\subsection{Equation of state of the cosmic string}

Without solving the equations of motion, it is possible to obtain the equation of state of the cosmic string. Indeed, we chose an ansatz for the fields where they only depend on $r$ and $\theta$. So, nothing in the configuration we are interested in can depend on internal string world sheet coordinates, here locally $z$ and $t$. The equation of state then gives \cite{Allys:2015yda,Hindmarsh:1994re}
\begin{equation}
U=T,
\end{equation}
where $U$ is the energy per unit length and $T$ is the tension of the string. This equation is the Nambu-Goto equation of state, which is Lorentz-invariant along the world sheet.

Thus, the only macroscopic parameter we will consider in the following is the energy per unit length, defined by
\begin{equation}
\label{DefU}
U=2\pi \int r \text{d}r \mathcal{L}.
\end{equation}

%%%%%%%%%%%%%%%%%%%%%%%%%%%%%%%%%%%%%%%%%%%%%%%%%%%%%%%%%%%%%%%%%
%%%%%%%%%%%%%%%%%%%%%%%%%%%%%%%%%%%%%%%%%%%%%%%%%%%%%%%%%%%%%%%%%
\section{Dimensionless model, perturbative study}
\label{PartScaling&Pert}

%%%%%%%%%%%%%%%%%%%%%%%%%%%%%%%%%%%%%%%%%%%%%%%%%%%%%%%%%%%%%%%%%
\subsection{Dimensionless model}
\label{PartScaling}

To work with a dimensionless model, we introduce the following new variables (denoted by a tilde)
\begin{equation}
\begin{matrix}
\displaystyle{ r = \frac{\tilde{r}}{\kappa M},} & \displaystyle{f = M \tilde{f},} & \displaystyle{S = \frac{m_{\mathbf{\Sigma}}}{\kappa} \tilde{S},}  \\[7pt]
  \displaystyle{a = \frac{m}{\lambda} \tilde{a},} & \displaystyle{b = \frac{m}{\lambda} \tilde{b},} & \displaystyle{p = \frac{m}{\lambda} \tilde{p},}\\[7pt]
  \displaystyle{F_X = \kappa M^2 \tilde{F_X},} & \displaystyle{V = \kappa^2 M^4 \tilde{V},} & \displaystyle{g^2 = \kappa^2 \tilde{g}^2}.
\end{matrix}
\end{equation}
Also, we introduce a set of dimensionless parameters, 
\begin{equation}
\begin{matrix}
\displaystyle{\alpha_1 = \frac{m}{\lambda M},} & \displaystyle{\alpha_2 = \frac{m_{\mathbf{\Sigma}}}{\kappa M},} \\[7pt]
\displaystyle{\alpha_3 = \frac{\eta}{\lambda},} & \displaystyle{\alpha_4 = \frac{\eta}{\kappa }},
\end{matrix}
\end{equation}
which, in addition to $\tilde{g}$, are the free parameters of the theory. As $\alpha_2$ and $g$ are real, while the other $\alpha_i$ are complex, the total parameter space of the model is of dimension 7 (the phases of $\alpha_1$, $\alpha_3$ and $\alpha_4$ are not independent).

Finally, the integrated Lagrangian over the radial coordinates $(r,\theta)$ gives
\begin{widetext}
\begin{multline}
\label{LagDensFull}
-\frac{L}{M^{2}}=2\pi \int \tilde{r} \text{d}\tilde{r} \left[  2\left(\tilde{f}^\prime\right)^2 +\frac{\text{Tr} \left({\tau_{\text{str}}}^2 \right) }{\tilde{g}^2}\frac{Q'^2}{2 \tilde{r}^2}
+  |\alpha_2|^2 \tilde{S}'^\ast \tilde{S}'
 + |\alpha_1|^2 \tilde{p}'^\ast \tilde{p}' + |\alpha_1|^2 \tilde{a}'^\ast \tilde{a}' 
 \right. \\ \left.
 +  |\alpha_1|^2 \tilde{b}'^\ast \tilde{b}'
+\frac{\tilde{f}^2Q^2}{\tilde{r}^2}+\tilde{V}(\tilde{\sigma},\tilde{a},\tilde{b},\tilde{p},\tilde{S}) \right],
\end{multline}
with 
\begin{multline}
\label{PotentialFull}
\tilde{V}={\left|\frac{\alpha_1^2\alpha_4}{\alpha_3}\right|}^2 \Bigg( 
 {\left|\tilde{p}+\frac{\tilde{b}^2}{6\sqrt{6}}+\frac{\alpha_3}{\alpha_1^2}\frac{\tilde{f}^2}{10\sqrt{6}} \right|}^2
+  {\left|\tilde{a}+\frac{\tilde{a}^2}{9\sqrt{2}}+\frac{\tilde{b}^2}{9\sqrt{2}}+\frac{\alpha_3}{\alpha_1^2}\frac{\tilde{f}^2}{10\sqrt{2}} \right|}^2
+ \left|\tilde{b} +\frac{\sqrt{2}\tilde{a}\tilde{b}}{9}+\frac{\tilde{b}\tilde{p}}{3\sqrt{6}}
 \right. \\ \left.
-\frac{\alpha_3}{\alpha_1^2}\frac{\tilde{f}^2}{10} \right|^2 \Bigg)
 + 2 {\left| \alpha_2 \right|}^2
\bigg| \tilde{f}
 +\frac{\alpha_1\alpha_4}{\alpha_2}\frac{\tilde{f}}{10}\left(\frac{\tilde{a}}{\sqrt{2}}-\tilde{b} + \frac{\tilde{p}}{\sqrt{10}} \right) +\tilde{S}\tilde{f}\bigg|^2 + {\left| \tilde{f}^2-1 \right|}^2.
\end{multline}
\end{widetext}

We can also write in a dimensionless form the sets of VEVs before and after the end of inflation, obtained as explained in Sec.~\ref{SSB&HI}. The analytical solutions being cumbersome, we give here only a Taylor expansion. The expansion parameter we consider is
\begin{equation}
\label{DefX}
x=\frac{\alpha_3}{\alpha_1^2}=\frac{\eta\lambda M^2}{m^2},
\end{equation}
which is often small in comparison with unity in such GUT models. Indeed, we expect the second step of symmetry breaking to appear at lower energy than the first step, and we mainly stay in a regime where the coupling constant are at most of order $1$. Note however that even if these expansions are convenient to describe the VEVs when $x\ll 1$, it is necessary to use the complete analytical solutions when it is not the case anymore.

For the first SSB scheme, going through $G'_2=3_C 2_L 2_R 1_{B-L}$, we have
\begin{equation}
\label{VEV0}
\left\{
\begin{array}{l}
\tilde{\sigma}_0=0,\\
\tilde{a}_0=-9 \sqrt{2},\\
\tilde{b}_0=0,\\
\tilde{p}_0=0,\\
\end{array}
\right.
\end{equation}
before the end of inflation, and
\begin{equation}
\label{VEV1}
\left\{
\begin{array}{l}
|\tilde{\sigma}_1|=1\\
\displaystyle{\tilde{a}_1=-9\sqrt{2}+\frac{x}{10\sqrt{2}}}\\
\displaystyle{\tilde{b}_1=-\frac{x}{10}}\\
\displaystyle{\tilde{p}_1=-\frac{x}{10\sqrt{6}}}\\
\displaystyle{\tilde{S}_1=-1+\frac{\alpha_1 \alpha_4}{\alpha_2}\left(\frac{9}{10} -\frac{x}{75} \right)}\\
\end{array}
\right.
\end{equation}
after the end of inflation. We recall that this previous set of VEVs defines the boundary conditions for the fields at infinity, as explained in Sec.~\ref{PartBoundaryConditions}. The set of VEV for the scheme going through $G'_1=3_C 2_L 1_R 1_{B-L}$ is given in the Appendix~\ref{AppendixVEVG2}. 

%%%%%%%%%%%%%%%%%%%%%%%%%%%%%%%%%%%%%%%%%%%%%%%%%%%%%%%%%%%%%%%%%
\subsection{Toy-model limit}
\label{PartOdGToyModel}

We will compare the results obtained to the abelian Higgs model. This toy model contains two scalar fields of opposite charges under a local U(1) gauge symmetry, $\Sigma$ and $\overline{\Sigma}$, and has for Lagrangian\footnote{
Note that if one wants to use a model where the kinetic part of the Lagrangian only contains a term in 
\begin{equation}
K_\Sigma = -(D_{\mu}\Sigma)(D^{\mu}\overline{\Sigma}),
\end{equation}
it it possible to make a link between both these toy-model considering $\tilde{f}=\sqrt{2} f$, $\tilde{\kappa}=\kappa/2$ and $\tilde{M} = \sqrt{2}M$, labeling with a tilde the expressions to use in the model with one single kinetic term. Indeed, the factor $2$ in the kinetic term will vanish, and the superpotential will stay identical.} \cite{Davis:1997bs,Aryal:1987sn,Peter:1992dw,Hindmarsh:1994re}
\begin{multline}
\label{LagToyModel}
\mathcal{L}=-(D_{\mu}\Sigma)^\dagger(D^{\mu}\Sigma)
-(D_{\mu}\overline{\Sigma})^\dagger(D^{\mu}\overline{\Sigma})\\
-\frac{1}{4} F_{\mu\nu}F^{\mu\nu}-\kappa^2 \left| \Sigma \overline{\Sigma}-M^2 \right| ^2.
\end{multline}
In order to describe it in a SUSY formalism, we have to introduce another field $S$, uncharged under the local symmetry, and use for the superpotential 
\begin{equation}
W=\kappa S \left( \overline{\Sigma} \Sigma -M^2\right).
\end{equation}
This yields the Lagrangian of Eq.~(\ref{LagToyModel}) when taking an ansatz where $S$ identically vanishes. In this toy model, the characteristic radii are $(\kappa M)^{-1}$ for $\Phi$ and $M^{-1}$ for the gauge field (see Appendix \ref{AppendixRCharac}).

This toy model can also be recovered from our realistic model, taking the limit $\eta \rightarrow 0$, and an ansatz where $\tilde{S}=-1$, and where the fields $a$, $b$ and $p$ identically take the value they have at infinity. Indeed, the fields associated to $\mathbf{\Phi}$ fully decouple to the string, $V_\Sigma$ and $V_{\overline{\Sigma}}$ vanish due to value of $\tilde{S}$ (see e.g. Eq.~(\ref{PotentialFull}), with $\alpha_4=0$), and $V_S$ reduces to the potential term of Eq.~(\ref{LagToyModel}). Note that we properly recover the toy-model case in this limit due to the normalization choice of Eq.~(\ref{SingletDefinition}). 

%%%%%%%%%%%%%%%%%%%%%%%%%%%%%%%%%%%%%%%%%%%%%%%%%%%%%%%%%%%%%%%%%
\subsection{Perturbative study}
\label{PartPerturbativeStudy}

We now perform a perturbative study of the condensation of the field $\mathbf{\Phi}$ in the string, in a certain range of parameter, following Ref.~\cite{Allys:2015yda}. For this purpose, considering the modifications of the fields $\tilde{a}$, $\tilde{b}$ and $\tilde{p}$ in a perturbative way when $\tilde{f}$ goes from $0$ to $1$, we obtain the characteristic scale of variation, e.g. for $\tilde{a}$, of [see Eq.~(\ref{PotentialFull})]\footnote{The small parameter we introduced in Eq.~(\ref{DefX}) in order to describe with a Taylor expansion the set of VEV at infinity appears in this characteristic scale of perturbation. This result is somehow natural, since we consider in both cases the static configurations which minimize the potential taking into account the constraint $\tilde{f}=0$ or $\tilde{f}=1$.}
\begin{equation}
\label{DefXPert}
\delta \tilde{a}_0 =\frac{\lambda \eta M^2}{ 10 \sqrt{2} m^2}=\frac{x}{10\sqrt{2}}.
\end{equation}
As we consider models where the end of inflation appears at a lower scale than the GUT scale, and with coupling constant often smaller than unity, this characteristic scale of variations is in most cases smaller than the scale of variation of $\tilde{f}$, which is $1$.  Finally, the ratio between the characteristic variation of $\tilde{a}$ and its dominant contribution at infinity $\tilde{a_0}$ gives [see Eqs.~(\ref{VEV0}) and~(\ref{VEV1})]
\begin{equation}
\frac{\delta \tilde{a}_0}{\tilde{a}_0} = \frac{\lambda \eta M^2}{ 180 m^2}=\frac{x}{180},
\end{equation}
which also legitimates the perturbation study.
This result is in accord with Ref.~\cite{Allys:2015yda}, which estimated this term to be of order $x/N$, with $N$ the characteristic dimension of the representations used. 

However, we could introduce a more precise estimate for the condensation of the fields coming from $\mathbf{\Phi}$ into the core of the string. Indeed, the value $\delta \tilde{a}_0$ we computed describes the differences between the configurations which minimize the potential in the center of the string and at infinity. But the actual value of this field $\tilde{a}$ results in a competition between the kinetic and the potential terms. In order to estimate this value, we can approximate at a linear order the contribution of this field to the Lagrangian close to the center of the string to 
\begin{equation}
\label{LagPert}
|\alpha_1|^2 \mathcal{L}_{\text{pert}} \simeq \left( \frac{\text{d} \delta\tilde{a}}{\text{d} \tilde{r}}\right)^2 + \left( \frac{m}{\kappa M } \right)^2 \left(\delta \tilde{a} -\delta\tilde{a}_0 \right)^2.
\end{equation}
Then, when $m/(\kappa M) \gg 1$, we obtain that $\delta \tilde{a} \simeq \delta \tilde{a}_0$. On the other hand, when $m/(\kappa M) \ll 1$, we obtain at dominant order that 
\begin{equation}
\delta \tilde{a}\simeq  \left( \frac{m}{\kappa M } \right) \delta\tilde{a}_0.
\end{equation}
To compute the kinetic term, we assume that the characteristic radius of the fields $a$, $b$ and $p$ is the same as the characteristic radius of $f$, i.e. $(\kappa M)^{-1}$, since these fields have a direct coupling in the Lagrangian with $f$ only, and not $Q$.

When the perturbative study is possible, we can now estimate the variation of the energy per unit length of the string due to the condensation of the additional fields into the string. For this purpose, it is sufficient to consider the kinetic contribution of the fields condensing in the core of the string, i.e. $\tilde{a}$, $\tilde{b}$ and $\tilde{p}$. Indeed, without this term, the potential would only play the role of a Lagrange multiplier for these fields, and it would not add any contribution to the energy of the string. An additional way to check this assumption is to verify that the kinetic and potential contributions of these fields to the Lagrangian density are similar.

These assumptions finally give a characteristic modification to the Lagrangian density of order 
\begin{equation}
\label{EqLArrach}
\delta \mathcal{L} \simeq |\alpha_1^{2}| \left( | \delta \tilde{a}|^2 + |\delta \tilde{b} |^2 +|\delta \tilde{p} |^2  \right),
\end{equation}
since in the Eq.~(\ref{LagDensFull}), the dimensionless radius used is $\tilde{r}$, in units of $(\kappa M)^{-1}$.
It yields, considering $U_0 \simeq M^2$ the energy per unit length of the toy-model string, and $\delta U$ the modification of the energy per unit length of the string due to the condensation of $\mathbf{\Phi}$ in the core of the string, 
\begin{equation}
\frac{\delta U}{U_0} \simeq \int \tilde{r} \text{d}\tilde{r} ~ \delta \mathcal{L} \simeq \delta \mathcal{L},
\end{equation}
which gives 
\begin{equation}
\label{PredictionUKappaPetit}
\frac{\delta U}{U_0} \simeq \frac{\eta^2 M^2}{60 m^2},
\end{equation}
when $m/(\kappa M) \gg 1$, and 
\begin{equation}
\label{PredictionUKappaGrand}
\frac{\delta U}{U_0} \simeq \frac{\eta^2 }{60 \kappa^2},
\end{equation}
when $m/(\kappa M) \ll 1$.
These evaluations of the modification of the energy per unit length of the string due to its realistic structure are relevant only when the perturbative approach is verified, i.e. when $x \ll 1$, and also when $\delta U / U_0$ does not approach unity. 

In Ref.~\cite{Allys:2015yda}, an estimate of the maximal modification of the energy per unit length from standard toy models was computed, considering the contribution from the scalar potential to the energy per unit length due to the condensation of the additional field in the core. This maximal estimate of $\delta U / U_0 \simeq \eta^2 /(N \kappa^2)$, with $N$ the characteristic dimension of the representations, is compatible with the results of Eqs.~(\ref{PredictionUKappaPetit}) and~(\ref{PredictionUKappaGrand}). The two results are very close in the second case, since as the additional fields barely condensate in the core of the string, the entire potential contribution used in Ref.~\cite{Allys:2015yda} has to be taken into account.

Note that we left aside at this point the contribution of the inflaton field $S$. It is possible to check that this contribution is of same order or lower than the contributions of the fields $a$, $b$ and $p$. On the other hand, it can be understood by the fact that $S$ has no characteristic scale, which is necessary for it in order to play the role of the inflaton.

%%%%%%%%%%%%%%%%%%%%%%%%%%%%%%%%%%%%%%%%%%%%%%%%%%%%%%%%%%%%%%%%%
%%%%%%%%%%%%%%%%%%%%%%%%%%%%%%%%%%%%%%%%%%%%%%%%%%%%%%%%%%%%%%%%%
\section{Numerical solution}
\label{PartNumericalSolution}

%%%%%%%%%%%%%%%%%%%%%%%%%%%%%%%%%%%%%%%%%%%%%%%%%%%%%%%%%%%%%%%%%
\subsection{Implementation in a real case}
\label{PartNumericalImplementation}

In order to simplify the model, we assume in the following that all the parameters, including $\lambda$ and $\eta$, are real. The numerical solution then reduces to a parameter space of $5$ dimensions, described e.g. by the $\alpha_i$ and $g$  which are real. Given this assumption, the boundary conditions for the fields at infinity given in Eq.~(\ref{VEV1}) are also real, since $x$ is real\footnote{When considering the analytical solutions, this result is valid only until $x\sim 22$.}.

With these real parameters, and since the potential is real, all the imaginary parts of the fields must appear at least in a quadratic form. It implies that the configuration where all the imaginary parts of the fields take an identically vanishing value is solution of the equations of motion. This solution is also compatible with the boundary conditions. We thus consider this ansatz from now on, and simplify the fields to real ones.

Moreover, the study of the minima of the potential with the constraint $\sigma=0$ has already been done since it is the configurations for the non vanishing VEVs before the end of inflation. Now, as this configuration is also real, it is compatible with the previous reality assumption done. Indeed, as discussed in \cite{Witten:1984eb}, the solutions for the fields in the core of the string result in a competition between the kinetic and the potential terms, and we expect fields to take values between the boundary conditions at infinity and the configuration which minimizes the potential in the center of the string.

\begin{figure}[t]
\begin{center}
\includegraphics[scale=0.44]{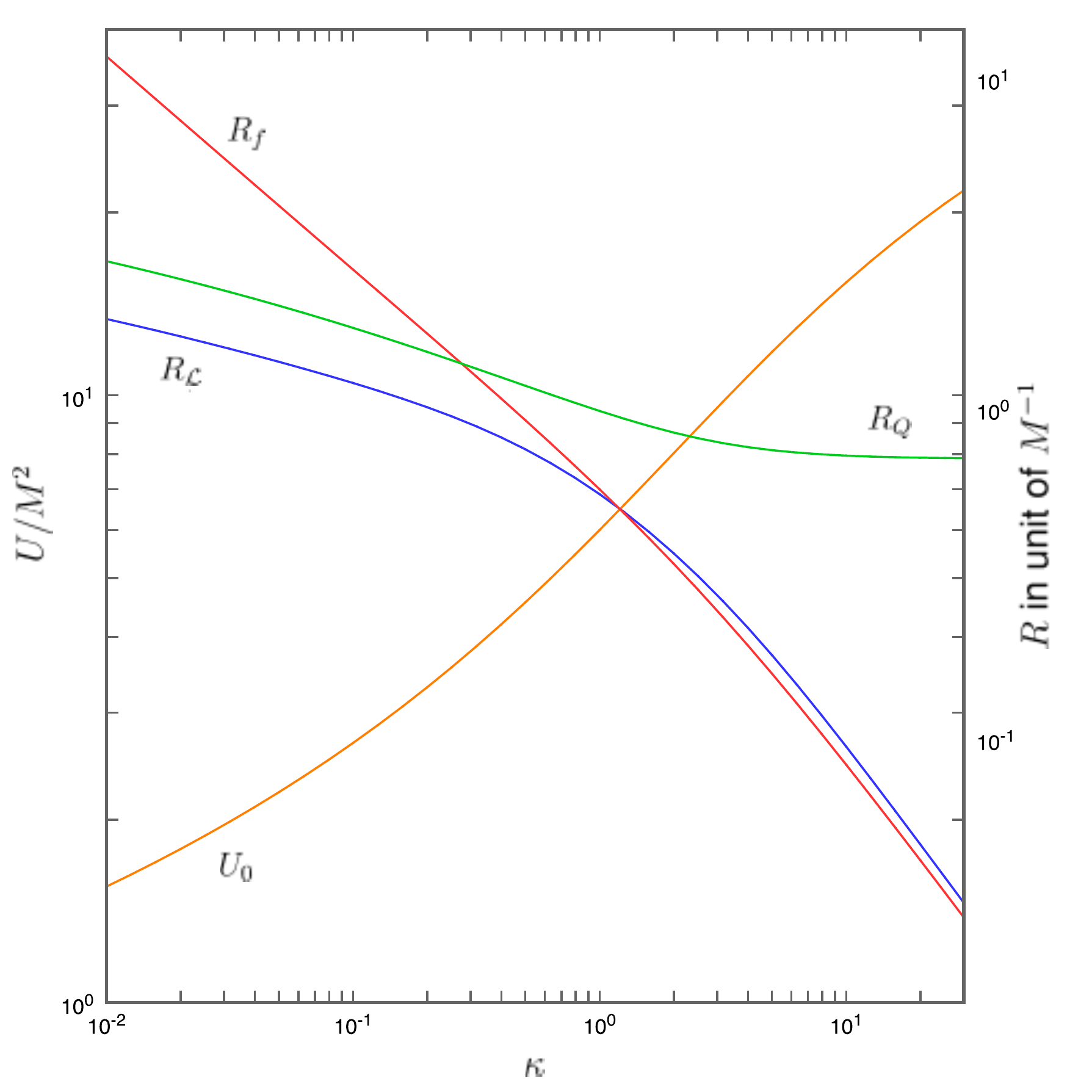}
\end{center}
 \caption{Energy per unit length $U$ and characteristic radius of the string ($R_{\mathcal{L}}$) and of the fields $f$ and $Q$ (resp. $R_f$ and $R_Q$) for the toy model limit ($\eta=0$). $U$ is in unit of $M^2$ and the radiuses in units of $M^{-1}$. }
 \label{Graph_eta=0}
\end{figure}

We suppose that this assumption will not change the results more than a small numerical factor. On the one hand, if the coupling constants $\lambda$ and $\eta$ were complex, there is no reason for their real and imaginary parts to be slightly different. Then, even if we were taking into account the imaginary part of the fields (using complex coupling constants), they would have similar equations of motion than their associated real part, and so would have comparable contributions.

In order to compute numerical solutions, we use a successively over-relaxed method to solve the equations of motion, after writing the whole model on a finite lattice (see e.g. Ref.~\cite{Adler:1983zh}). For this purpose, we convert the integral to a finite range one, introducing a variable $\rho = \tan r$. The numerical solution solves the equation of motion by minimizing the Lagrangian, which reduces to an algebraic function of the fields after being written on the lattice. For this purpose, we use successive Newton iterations, introducing an over-relaxation parameter $\omega$. For example, computing the root of an equation $f(x)=0$, the value of $x$ at the iteration $n+1$ thus gives
\begin{equation}
x^{n+1} = x^n - \omega \frac{f(x^n)}{f^\prime (x^n)}.
\end{equation}
This whole method is called the successively over-relaxed Gauss-Seidel iteration. For this kind of problem, keeping $0<\omega<2$ provides that the Lagrangian monotonously decreases at each step, the convergence being exponential for values of $\omega$ close to $2$. The precision of the result can be evaluated and is of the order of the square root of the modification of the Lagrangian in one step of over-relaxation at the end of the implementation, around $10^{-8}$ in our case. Also, we obtain the same numerical values with an accuracy better than $10^{-8}$ when either imposing the values of the fields at infinity or only a vanishing condition for the derivative of the fields. The lattices used contain $2000$ nodes. A scale factor on the radius is used in order to adapt the characteristic size of this lattice to the one of the string.

\begin{figure}[t]
\begin{center}
\includegraphics[scale=0.44]{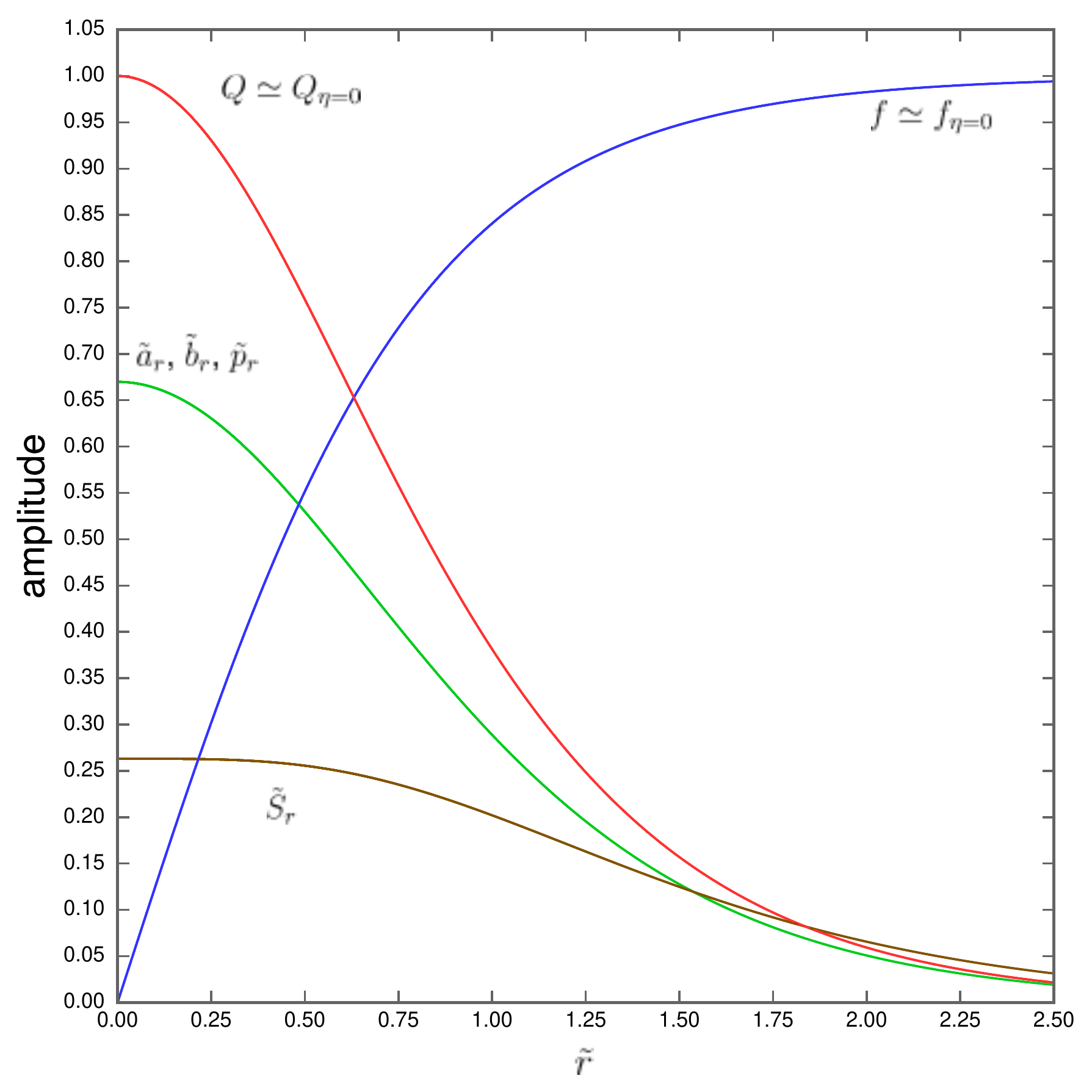}
\end{center}
 \caption{Structure of the cosmic string for $m/M=2$ and $\kappa=\lambda=\eta=1$. The values of $f$ and $Q$ cannot be distinguished between the case $\eta=1$ and $\eta=0$. The curves of $\tilde{a}_r$, $\tilde{b}_r$ and $\tilde{p}_r$ cannot either be distinguished.}
 \label{Graph_Structure1}
\end{figure}

\begin{figure}[t]
\begin{center}
\includegraphics[scale=0.218]{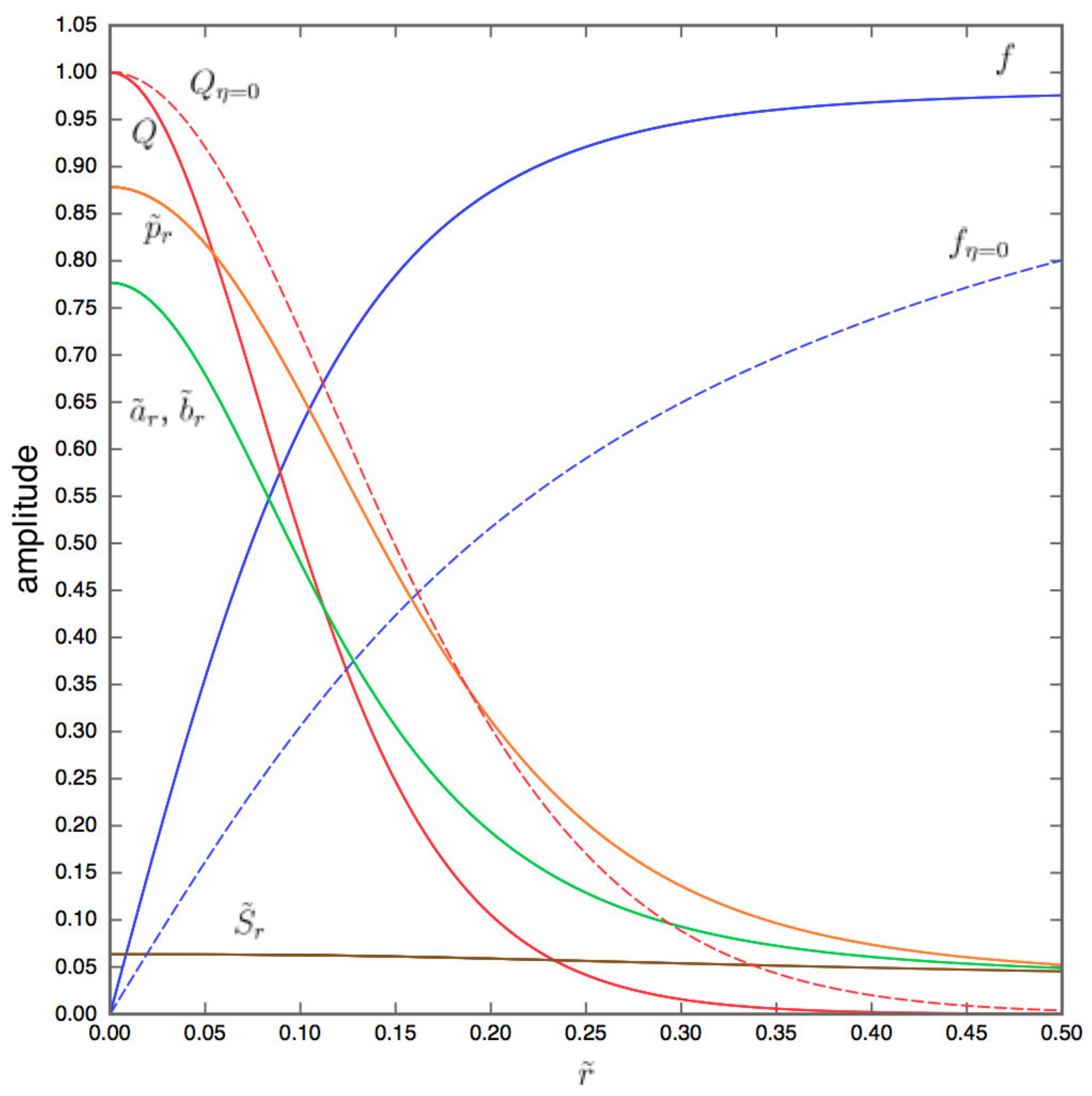}
\end{center}
 \caption{Structure of the cosmic string for $m/M=2$, $\kappa=0.1$ $\lambda=1$ and $\eta=10$. In dashed lines are the values of $f$ and $Q$ in the toy model limit $\eta=0$. The curves of $\tilde{a}_r$ and $\tilde{b}_r$ cannot be distinguished.}
 \label{Graph_Structure2}
\end{figure}

%%%%%%%%%%%%%%%%%%%%%%%%%%%%%%%%%%%%%%%%%%%%%%%%%%%%%%%%%%%%%%%%%
\subsection{Range of parameters, high-coupling limit}
\label{PartRangeParameters}

Let us now discuss the range of parameter chosen for the numerical solution. This part of the investigation permits to test the range of parameters for which the perturbative expansion is not valid, i.e. a high coupling limit for the additional fields condensing in the string. This limit is mainly achieved for large $M/m$ and $\eta$ and small $\kappa$, which we will consider.

The two masses $m$ and $m_{\Sigma}$ are set to be equal, presumably around $E_{\text{GUT}}\sim 10^{16}$GeV, and the energy scale $M$, characteristic of the end of inflation, takes values between $m$ and $m/20$. It is not possible to go to values of M/m larger than 1, since it is not compatible with the cosmological evolution we assumed (i.e. the order of the phase transition). We consider values of $\eta$ up to 10, which is already a high coupling in what concerns the GUT sector. Following the discussions of Ref.~\cite{Jeannerot:1995yn}, we take for $\kappa$ values between $0.01$ and $30$. The upper limit taken for $\lambda$ is one, but as it has a very small impact on the string, we leave it aside in most of the results presented here. The limit where the coupling constants and the mass ratio go to zero are well defined and described most in the case by the perturbative expansion, as discussed below. We considered values of coupling constants $\lambda$ and $\eta$ down to $10^{-2}$. For all the solutions, we take $g=1$, and a winding number unity. We also take $\text{Tr}( {\tau_{\text{str}}}^2)=2/5$ \cite{Ma:1992ky}.

Around 2000 different sets of parameters have been examined in the whole range discussed above.

%%%%%%%%%%%%%%%%%%%%%%%%%%%%%%%%%%%%%%%%%%%%%%%%%%%%%%%%%%%%%%%%%%
\subsection{Toy model limit}
\label{PartDefRadius}

To describe the structure of the string, we define different characteristic radiuses, related to the string itself, or to a field. When normalizing a field in order for it to have the value of $1$ in the center of the string, and $0$ at infinity, the characteristic radius of this field verifies $\phi(r_\phi)=0.40$. The characteristic radius of the string verifies the same property for the Lagrangian density $\mathcal{L}$.

The microscopic structure of the toy model string, i.e. the fields as functions of the radial coordinate $\tilde{r}$, can be found in Figs. \ref{Graph_Structure1} and \ref{Graph_Structure2}. Still for the toy-model case, we give in Fig.~\ref{Graph_eta=0} the value of the energy per unit length in unit of $M^2$, as well as the characteristic radius in units of $M^{-1}$ of $f$, $Q$ and of the Lagrangian density of the string, both as functions of $\kappa$. We properly recover the results of Sec.~\ref{AppendixRCharac}. The energy per unit length of this toy-model string verifies $U \simeq (1-20)M^2$, a common result found in the literature \cite{Kibble:1982ae,Ma:1992ky,Davis:1996sp,Davis:1997bs,Ferreira:2002mg,Morris:1997ua,Davis:1997ny,Adler:1983zh,Peter:1992dw}. From now on, we will compare the results obtained in the realistic implementation of the cosmic string to these particular numerical values, taking the associated toy-model for which all the parameters are the same but $\eta$ goes to zero.

%%%%%%%%%%%%%%%%%%%%%%%%%%%%%%%%%%%%%%%%%%%%%%%%%%%%%%%%%%%%%%%%%%
\subsection{Microscopic structure of the realistic string}

\begin{figure}[t]
\begin{center}
\includegraphics[scale=0.44]{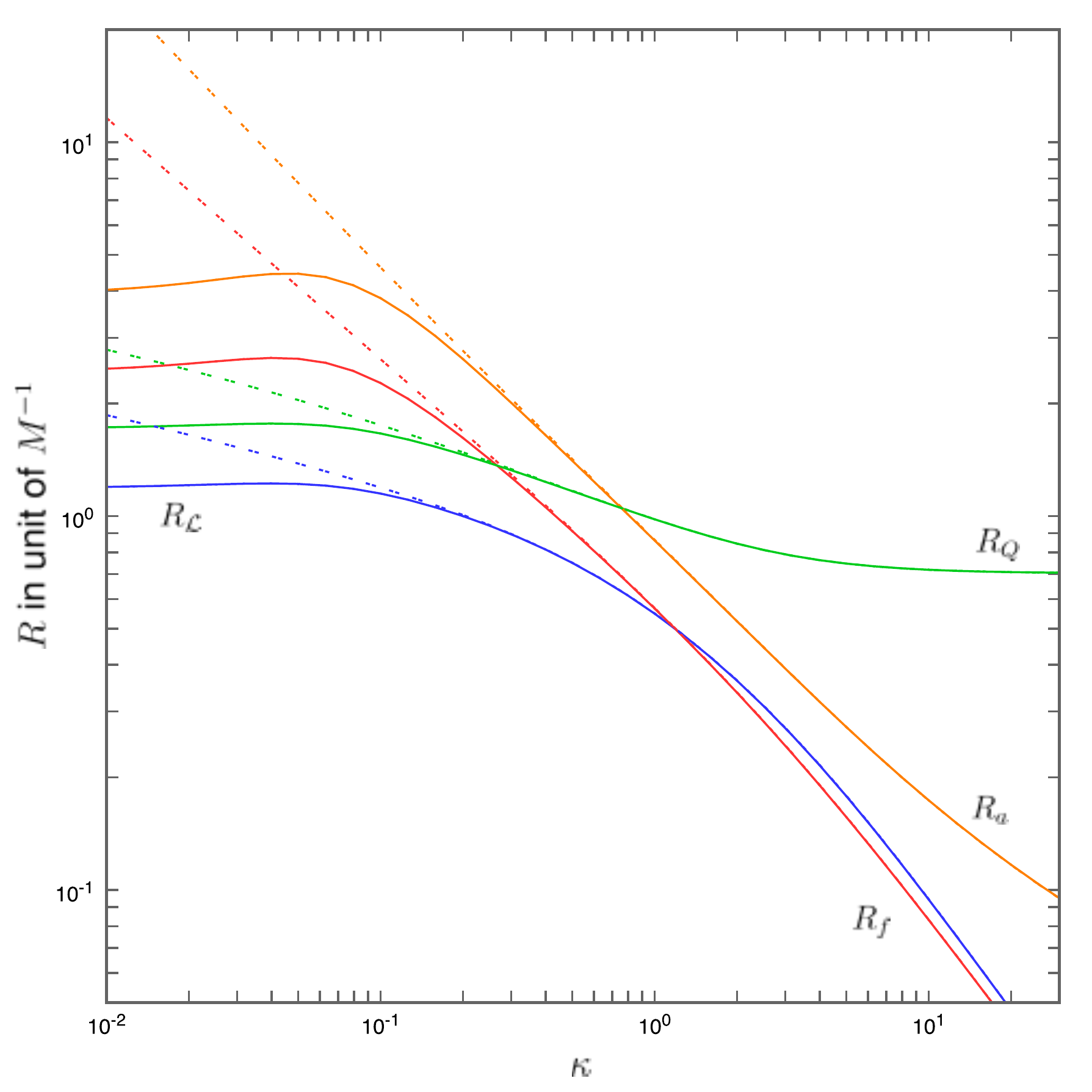}
\end{center}
 \caption{Characteristic radius of the string $R_{\mathcal{L}}$, and of the fields $f$, $Q$ and $a$ ($R_f$, $R_Q$ and $R_a$) as functions of $\kappa$, obtained for $\eta=10$ in solid lines, and for $\eta=0.1$ in dashed lines. In both case, $m/M=10$ and $\lambda=1$.}
 \label{Graph_Rayons_EtaFixe}
\end{figure}

\begin{figure}[t]
\begin{center}
\includegraphics[scale=0.44]{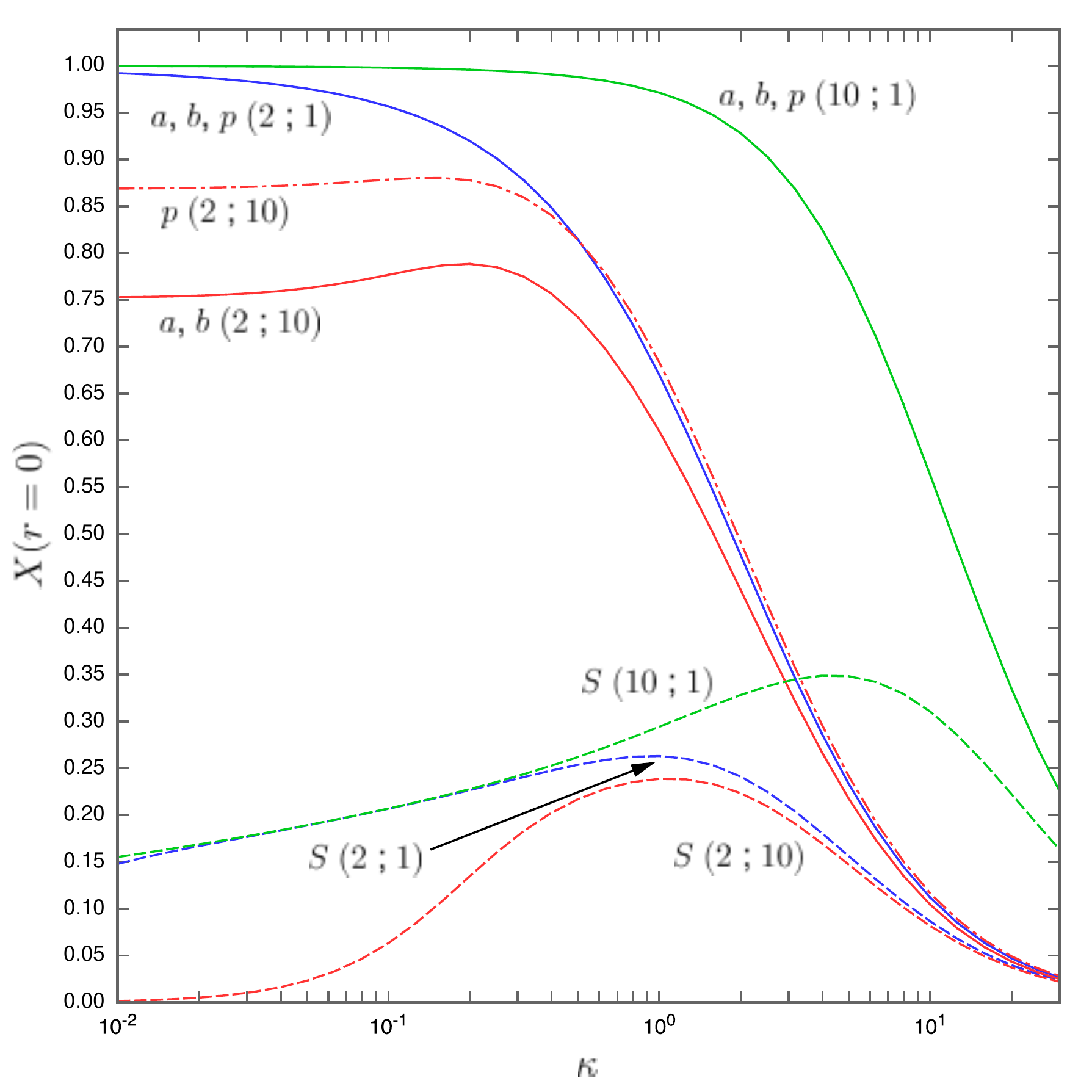}
\end{center}
 \caption{Value at $\tilde{r}=0$ of $\tilde{a}_r$, $\tilde{b}_r$, $\tilde{p}_r$ and $\tilde{S}_r$, as functions of $\kappa$. The different sets of parameters are labeled by $(m/M$ ; $\eta)$, with $m/M=2$ and $\eta=1$ in blue, $m/M=2$ and $\eta=10$ in red, and $m/M=10$ and $\eta=1$ in green, while $\lambda=1$ in all the configurations. For the blue and green curves, $a$, $b$ and $p$ cannot be distinguished and are in solid lines. For the red curves, $a$ and $b$ cannot be distinguished and are in solid lines, while $p$ is in dashed-pointed line. In all cases, $S$ is plotted in dashed lines.}
 \label{Graph_X_0}
\end{figure}

In order to show a graphic representation of the fields $\tilde{a}$, $\tilde{b}$, $\tilde{p}$ and $\tilde{S}$, we normalize them between $0$ and $1$, taking $0$ for their values at infinity, and $1$ for the static configuration which minimizes the potential at the center of the string, i.e. for $f=0$. For the inflaton field, we choose the configuration which minimizes the potential in the center of the string for small but not vanishing values of $\tilde{f}$. Indeed, when $\tilde{f}=0$, the inflaton field has a flat potential at tree level (see discussion of Sec.~\ref{SSB&HI}). We denote these normalized fields $\tilde{a}_r$, $\tilde{b}_r$, $\tilde{p}_r$ and $\tilde{S}_r$. This additional normalization for these fields uses the characteristic scale of variation of the problem: the value of these fields in the center of the string tends to $1$ if the potential term is dominant, and to zero if the kinetic term is dominant.

%\begin{figure}[t]
%\begin{center}
%\includegraphics[scale=0.44]{Graphiques/densiteL/densiteL.pdf}
%\end{center}
% \caption{Lagrangian density in unit of $M^2$ as a function of the radial coordinate $\tilde{r}$. In red, the configurations where $m/M=2$ and $\kappa=\lambda=1$, in blue where $m/M=2$, $\kappa=0.1$ and $\lambda=1$. For the case $\kappa=1$, the curves with $\eta=1$ and $\eta=0$ cannot be distinguished.}
% \label{Graph_densiteL}
%\end{figure}

Two microscopic structures of cosmic strings are given in Fig.~\ref{Graph_Structure1} and Fig.~\ref{Graph_Structure2}. Note that we plot in dashed lines the values of the fields $f$ and $Q$ in the toy-model limit of these strings. In the first configuration plotted in Fig.~\ref{Graph_Structure1}, $f$ and $Q$ are very close to their toy-model values, and cannot be distinguished from them in the graphic. The values of the fields $\tilde{a}_r$, $\tilde{b}_r$ and $\tilde{p}_r$ are also very close, and the associated curves merge together. In the second configuration, given in Fig.~\ref{Graph_Structure1}, the realistic structure of the string causes its radius to lower. In this configuration, the fields $\tilde{a}_r$ and $\tilde{b}_r$ are still very close in values, and cannot be distinguished in the graph. Note that in the second figure, the fields $\tilde{f}$, $\tilde{a}_r$, $\tilde{b}_r$, $\tilde{p}_r$ and $\tilde{S}$ properly converge at high radii. In this range, they recover the behavior they have for the same set of parameters but $\eta=1$.

In both configurations, the value of the perturbation parameter of order $x/10$ (see Eq.~(\ref{DefXPert}) and the subsequent discussion) is respectively $1/40$ and $1/4$, which is in agreement with the results observed, i.e. minor modifications from the toy model in Fig.~\ref{Graph_Structure1}, and sizable modifications in Fig.~\ref{Graph_Structure2}.

The similar behavior of the fields $\tilde{a}_r$, $\tilde{b}_r$ and $\tilde{p}_r$ can also be understood with the perturbative approach. In this limit, only the linear terms in these fields can be considered in $F_a$, $F_b$ and $F_p$, see Eqs.~(\ref{EqFp}-\ref{EqFb}) or (\ref{LagDensFull}), and they all have the same coupling with $\tilde{f}$ only. Thus, these fields have very close values after normalization. In Fig.~\ref{Graph_Structure2}, one can note that the fields $\tilde{a}_r$ and $\tilde{b}_r$ still have similar values, which is not the case anymore for the field $\tilde{p}_r$. This difference can be explained by the fact that $\tilde{a}_r$ and $\tilde{b}_r$ have quadratic terms in the $F$-terms, whereas $\tilde{p}_r$ only has linear terms. Note that in all these cases, one can verify that the kinetic and potential contributions to the Lagrangian density are of the same order.

\begin{figure}[t]
\begin{center}
\includegraphics[scale=0.44]{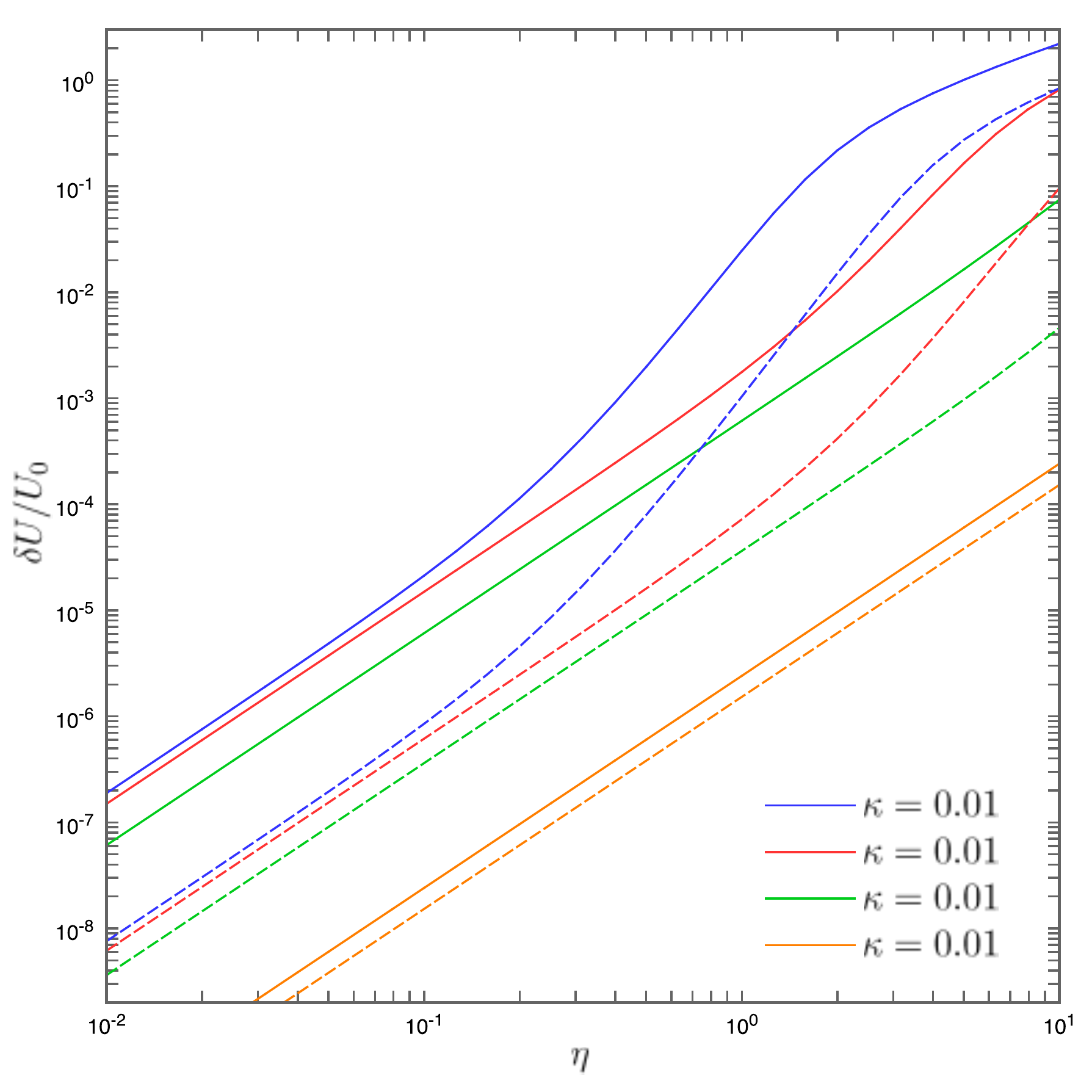}
\end{center}
 \caption{Normalized modifications of the energy per unit length from the toy model values, $\delta U / U_0$ as functions of $\eta$, for $\lambda=1$, and for various values of $\kappa$ and $m/M$. The constant lines are for $m/M=2$, and the dashed lines for $m/M=10$.}
 \label{Graph_GlobU_KappaFixe}
\end{figure}

\begin{figure}[t]
\begin{center}
\includegraphics[scale=0.44]{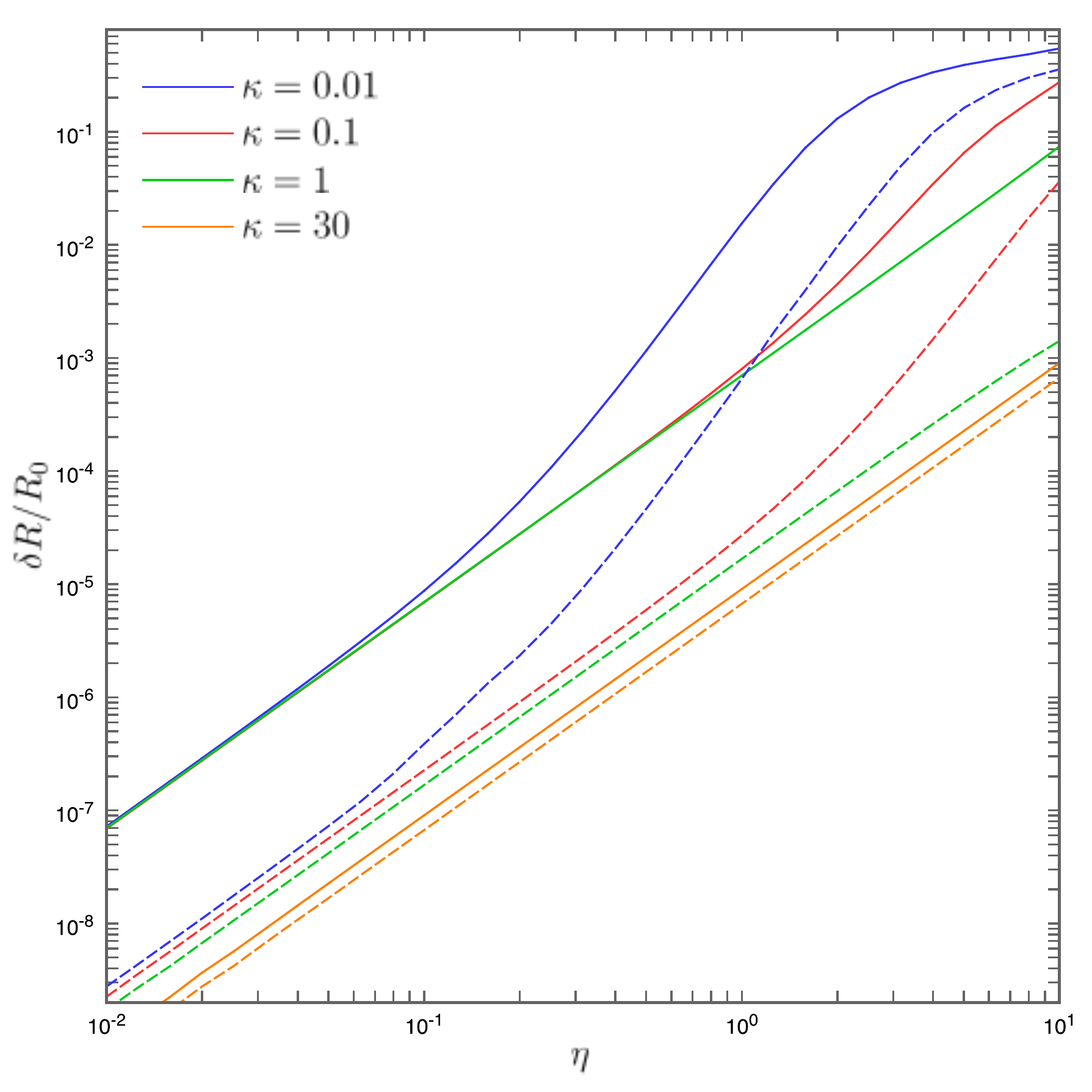}
\end{center}
 \caption{Normalized modifications of the radius of the string from the toy model values, $\delta R_{\mathcal{L}} / R_{\mathcal{L},0}$ as functions of $\eta$, for $\lambda=1$, and for various values of $\kappa$ and $m/M$. The constant lines are for $m/M=2$, and the dashed lines for $m/M=10$.}
 \label{Graph_GlobR_KappaFixe}
\end{figure}

In order to study the general shape of the microscopic structure, the radii of the string and of different fields are given in Fig.~\ref{Graph_Rayons_EtaFixe}. As before, two different behaviors appear in this figure. For $\eta=0.1$, the modification from the toy-model due to the realistic structure of the string is negligible (see Fig.~\ref{Graph_eta=0}). For $\eta=10$, important modifications appear for small values of $\kappa$. In this last limit, the perturbative approach considered before is not valid anymore. We also see that the coupling of the string-forming Higgs fields with other fields of higher energy tightens the radius of the string. 

To describe the condensation of the fields $\tilde{a}$, $\tilde{b}$, $\tilde{p}$ and $\tilde{S}$ in the string, we give in Fig.~\ref{Graph_X_0} the values of $\tilde{a}_r$, $\tilde{b}_r$, $\tilde{p}_r$ and $\tilde{S}_r$ in the center of the string, i.e. at $\tilde{r}=0$, for different configurations. When the perturbative study is possible, we recover the behavior discussed in Sec.~\ref{PartPerturbativeStudy} for the fields  $\tilde{a}$, $\tilde{b}$ and $\tilde{p}$, with a constant value around $1$ for $m/(\kappa M) \gg 1$, and a limit in $m/(\kappa M)$ for $m/(\kappa M) \ll1$. Note that a numerical coefficient appears in the asymptotic behavior, coming from the rough estimate done in Eq.~(\ref{EqLArrach}), and of the form 
\begin{equation}
\label{DefBeta}
{a}_r (0) \sim \frac{\beta m}{\kappa M},
\end{equation}
with $\beta$ of order $\simeq 0.5$.

Such a reasoning for the inflaton field $S$ is not as simple, since its kinetic and potential terms in dimensionless forms have similar prefactors in the Lagrangian, see Eqs.~(\ref{LagDensFull}) and~(\ref{PotentialFull}). It could explain nevertheless why the inflaton field never fully condensates in the string, i.e. with values of $\tilde{S}_r$ close to unity. The inflaton also has an asymptotic behavior in $m/(\kappa M)$ for $m/(\kappa M) \ll1$.

In the previous results, we left aside the study of the microscopic structure as a function of the parameter $\lambda$. Indeed, varying this parameter from $0.01$ to $1$ in the previous configurations only has a minor impact on the model, and barely modifies the graphic results obtained.

\begin{figure}[t]
\begin{center}
\includegraphics[scale=0.44]{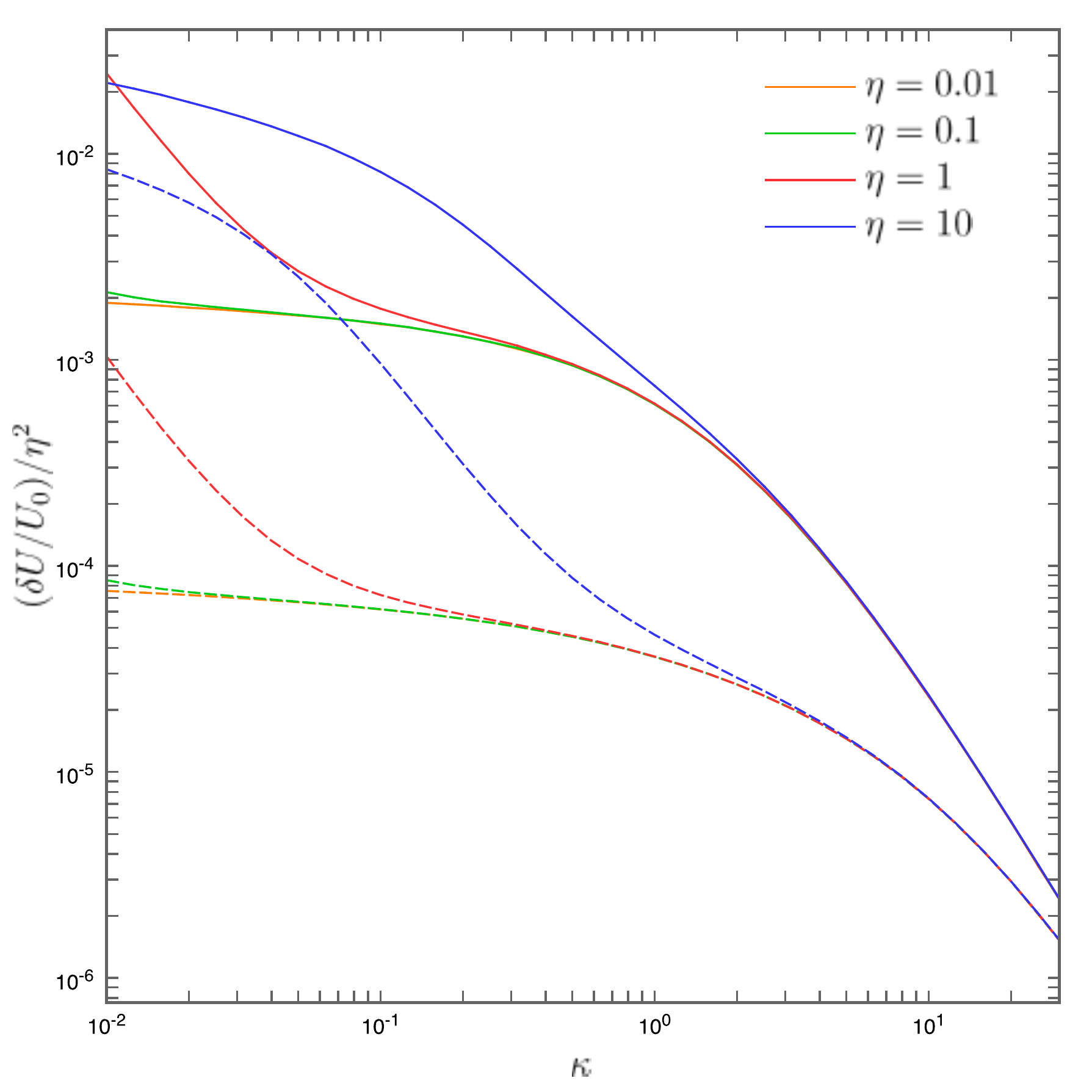}
\end{center}
 \caption{Normalized modification of the energy per unit length from the toy-model values $\delta U / U_0$, divided by $\eta^2$, as functions of $\kappa$, for various values of $\eta$ and $m/M$, and for $\lambda=1$. The constant lines are for $m/M=2$, and the dashed lines for $m/M=10$.}
 \label{Graph_GlobU_EtaFixe_Eta2}
\end{figure}

\begin{figure}[t]
\begin{center}
\includegraphics[scale=0.43]{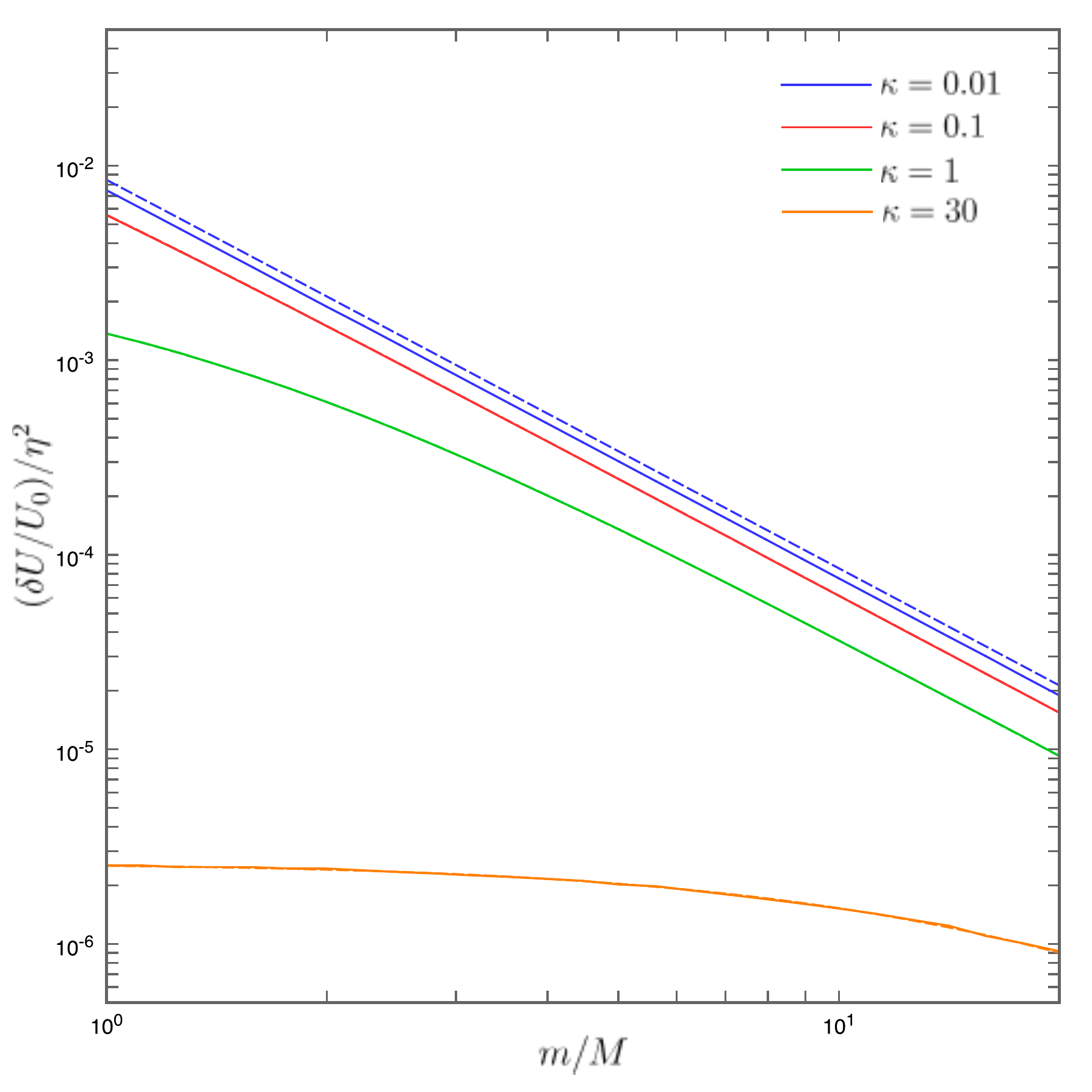}
\end{center}
 \caption{Normalized modifications of the energy per unit length from the toy-model values, $\delta U / U_0$, divided by $\eta^2$, as functions of $m/M$, for various values of $\eta$ and $\kappa$, and for $\lambda=1$. The constant lines are for $\eta=0.01$, and the dashed lines for $\eta=0.1$. These curves can be distinguished for $\kappa=0.01$ only.}
 \label{Graph_GlobU_xVariable_Eta2}
\end{figure}

%%%%%%%%%%%%%%%%%%%%%%%%%%%%%%%%%%%%%%%%%%%%%%%%%%%%%%%%%%%%%%%%%%
\subsection{Macroscopic properties of the realistic string}

Let us turn now to the modifications of the energy per unit length and the radius of the strings, as functions of the different GUT parameters. Fig.~\ref{Graph_GlobU_KappaFixe} shows the modifications of the energy per unit length from the toy-model limit, i.e. $(U-U_0)/U_0$ as a function of $\eta$ where $U_0$ is the energy of the toy-model associated to the same GUT parameters (described in Sec.~\ref{PartOdGToyModel}) which can be found in Sec.~\ref{PartDefRadius}. Similarly, Fig.~\ref{Graph_GlobR_KappaFixe} shows the evolution of $(R_{\mathcal{L}}-R_{\mathcal{L},0})/R_{\mathcal{L},0}$, as a function of $\eta$, where $R_{\mathcal{L},0}$ is defined similarly to $U_0$ and can be found in Sec.~\ref{PartDefRadius}.
The results obtained in Sec.~\ref{PartPerturbativeStudy} are verified, with a behavior proportional to $\eta^2$ when the perturbative study is valid. 

In addition, we see that the modifications of the radius of the string are close to the modifications of the energy per unit length. This result is verified for the whole range of parameters studied here. It is not particularly relevant to discuss the small differences between these both curves, since there is some arbitrariness in the definition of the radii. Also, an inflection point appears in both these figures when the radius of the string is modified more than a few percent. At this point, the decrease of the radius of the string seems to partially balance the augmentation of the energy due to the condensation of the additional fields in the core of the string. It can also be explained by the fact that the parameter $m$ becomes important in the description of the string, parameter which is associated with a typical length $\sim m^{-1}$ smaller than $M^{-1}$ the typical length of the toy-model string.

The dependencies in $\kappa$ and $m/M$ of the modification of the energy per unit length are plotted in~\ref{Graph_GlobU_EtaFixe_Eta2} and \ref{Graph_GlobU_xVariable_Eta2} respectively. In both of these figures, the modification of energy per unit length is divided by $\eta^2$. We recover the results obtained in the perturbative study, see Eqs.~(\ref{PredictionUKappaPetit}) and~(\ref{PredictionUKappaGrand}), i.e. $\delta U/U_0 \simeq \eta^2 M^2/(60 m^2)$ for large $m/(\kappa M)$, and $\delta U/U_0 \simeq \eta^2/(60 \kappa^2)$ for small $m/(\kappa M)$. Non perturbative sets of parameters are also presented in Fig.~\ref{Graph_GlobU_EtaFixe_Eta2}. 

Finally, it is possible to evaluate the numerical parameter which appears in Eqs.~(\ref{PredictionUKappaPetit}) and~(\ref{PredictionUKappaGrand}), and was approximate to $1/60$. For the behavior at small $\kappa$, we obtain values close to $1/120$ with is in good agreement with the perturbative result. In the case of large $\kappa$, we obtain results around $1/500$. However, it is also in good agreement with the perturbative results after taking into account the additional parameter $\beta$ defined in Eq.~(\ref{DefBeta}).

In what concerns the macroscopic properties of the string, we also left aside the variations with respect to the parameter $\lambda$. Indeed, varying this parameter from $0.01$ to $1$ also barely modifies the graphic results given here.

%%%%%%%%%%%%%%%%%%%%%%%%%%%%%%%%%%%%%%%%%%%%%%%%%%%%%%%%%%%%%%%%%
%%%%%%%%%%%%%%%%%%%%%%%%%%%%%%%%%%%%%%%%%%%%%%%%%%%%%%%%%%%%%%%%%
\section{Conclusion and discussion}

In this paper, we performed a complete study of the realistic structure of cosmic strings forming in a given SO(10) SUSY GUT. Writing this GUT with tensorial representations, we showed that it was possible to simplify this study to a few complex functions describing the dynamics of the sub-representations which are singlet under the SM symmetry. We gave a full ansatz for a string in this context, and performed a perturbative study of the model obtained. We then presented numerical solutions of the string structure, and discussed their microscopic and macroscopic properties, which involve a rich phenomenology.

The numerical results showed that the modification of the energy per unit length from standard toy model strings is modified by a factor slightly higher than unity in the high coupling limit, which is already important with regards to CMB constraints. Note that we tried to investigate the largest possible available range of parameters for which the modifications of the macroscopic properties of the strings is sizable, see discussion of Sec.~\ref{PartRangeParameters}. Getting stronger modifications would require more extreme values of the parameters, values which would then be questionable for the reasons discussed above. Whatever, it shows that in this high-coupling limit, the simplest toy-models seem not to be appropriate models to describe the macroscopic properties of the strings, and corrections due to their realistic structure should be taken into account. In addition, the contribution of the inflaton field to the macroscopic characteristics of the strings appears to be negligible. It is an additional indication that including or not the inflaton in the GUT field content has no major impact on the cosmic strings properties.

The perturbative expansion is in very good agreement with numerical solutions in the wide range of parameters where this approach is possible. The precise microscopic structure gives several different criteria to test the perturbative results, and strengthens their relevance. These results show that a perturbative expansion of the realistic structure of cosmic strings around the toy-mode one is a reliable method to study them. It is particularly useful since such studies are often permitted, especially with the wide numerical factors coming from the large dimensional representations used in GUT. It enhances the result that the modifications of the macroscopic properties of so-called single-field strings, i.e. with no coupling of the form $\beta \mathbf{\Phi}\mathbf{\Sigma}\overline{\mathbf{\Sigma}}$ between the singlets of the SM in the superpotential, become sizable in a very high coupling limit~\cite{Allys:2015yda}, as it is the case here. Also, the present work shows that when a perturbative study is not possible anymore, the modifications of the structure and properties of the strings can become important. It strengthens the idea that in the case of the so-called many-fields strings, i.e. with a coupling of the form $\beta \mathbf{\Phi}\mathbf{\Sigma}\overline{\mathbf{\Sigma}}$~\cite{Allys:2015yda}, a complete study is necessary and should be done in the future, since no perturbative discussion is possible in most of the range of parameters.

In this work, we left aside precise considerations about the stability of the ansatz we used. We should consider this in more details in the future. Other properties of the strings should also be studied in the realistic GUT context considered here, such as the existence of bosonic currents in the core of the string \cite{Witten:1984eb,Peter:1992dw,Peter:1992nz,Peter:1993tm,Morris:1995wd}. Moreover, and as we work in a supersymmetric framework, their superpartner could carry fermionic currents through their zero modes \cite{Witten:1984eb,Jackiw:1981ee,Weinberg:1981eu,Davis:1995kk,Ringeval:2000kz,Peter:2000sw}. The effect of this complete structure to the processes of intercommutation \cite{Shellard:1988ki,Laguna:1990it,Matzner:1988ky,Moriarty:1988qs,Moriarty:1988em,Shellard:1987bv} should also be investigated, keeping in mind that modifications of such properties of the cosmic strings could affect the network evolution, and thus the observational consequences on the CMB \cite{Kibble:1976sj,Kibble:1980mv,Copeland:1991kz,Austin:1993rg,Martins:1996jp,Martins:2000cs}. Finally, the possibility of formation of non-abelian strings could also be studied in this context, and gives some interesting phenomenology\cite{Aryal:1987sn,Ma:1992ky,Davis:1996sp}.

%%%%%%%%%%%%%%%%%%%%%%%%%%%%%%%%%%%%%%%%%%%%%%%%%%%%%%%%%%%%%%%%%%

\subsection*{Acknowledgment}

I wish to thank P. Peter and M. Sakellariadou for many valuable discussions and suggestions, and also for a critical reading of the manuscript, which permitted important improvements on the present paper. I also thank J. Allys and J.-B. Fouvry for useful advice on numerical solutions and data processing.

%%%%%%%%%%%%%%%%%%%%%%%%%%%%%%%%%%%%%%%%%%%%%%%%%%%%%%%%%%%%%%%%%
%%%%%%%%%%%%%%%%%%%%%%%%%%%%%%%%%%%%%%%%%%%%%%%%%%%%%%%%%%%%%%%%%
%%%%%%%%%%%%%%%%%%%%%%%%%%%%%%%%%%%%%%%%%%%%%%%%%%%%%%%%%%%%%%%%%
%%%%%%%%%%%%%%%%%%%%%%%%%%%%%%%%%%%%%%%%%%%%%%%%%%%%%%%%%%%%%%%%%
%%%%%%%%%%%%%%%%%%%%%%%%%%%%%%%%%%%%%%%%%%%%%%%%%%%%%%%%%%%%%%%%%
%%%%%%%%%%%%%%%%%%%%%%%%%%%%%%%%%%%%%%%%%%%%%%%%%%%%%%%%%%%%%%%%%
%%%%%%%%%%%%%%%%%%%%%%%%%%%%%%%%%%%%%%%%%%%%%%%%%%%%%%%%%%%%%%%%%
%%%%%%%%%%%%%%%%%%%%%%%%%%%%%%%%%%%%%%%%%%%%%%%%%%%%%%%%%%%%%%%%%

\appendix
\section{Intermediate calculations}
\label{AppendixA}
%%%%%%%%%%%%%%%%%%%%%%%%%%%%%%%%%%%%%%%%%%%%%%%%%%%%%%%%%%%%%%%%%
\subsection{Computation of the derivatives}
\label{AppendixDerivativeComputation}

We present here how to take derivative with respect to the fields of the GUT in a tensor formulation. We take as an example the computation of $(\partial \{\mathbf{\Phi}\mathbf{\Sigma}\overline{\mathbf{\Sigma}}\} )/ (\partial \mathbf{\Sigma})$. Computing this derivative, we have to take into account that the different components of the multiplet $\mathbf{\Sigma}$ are not independent. 

Starting from
\begin{equation}
\mathbf{\Phi}\mathbf{\Sigma}\overline{\mathbf{\Sigma}} = \Sigma_{ijklm}\Phi_{ij\alpha\beta}\overline{\Sigma}_{\alpha\beta klm},
\end{equation}
we can use the fact that $\mathbf{\Sigma}$ is self-dual, see Eq. (\ref{self-dual}), to write 
\begin{equation}
 = \frac{1}{2}\left(\Sigma_{ijklm}+\frac{i}{5!}\epsilon_{ijklmabcde}\Sigma_{abcde}\right)\Phi_{ij\alpha\beta}\overline{\Sigma}_{\alpha\beta klm}.
\end{equation}
Now, as $\mathbf{\Sigma}$ is totally antisymmetric, we can write
\begin{equation}
 = \frac{1}{2}\left(\Sigma_{[ijklm]}+\frac{i}{5!}\epsilon_{ijklmabcde}\Sigma_{abcde}\right)\Phi_{ij\alpha\beta}\overline{\Sigma}_{\alpha\beta klm},
\end{equation}
where the antisymmetrization is on all the indices of $\mathbf{\Sigma}$. There is no need to make it appear explicitly in the Levi-Civita symbol since it is already totally antisymmetric. Then, we wrote all the component of $\mathbf{\Sigma}$ which are not independent. It gives after a relabeling of the indices :
\begin{multline}
\mathbf{\Phi}\mathbf{\Sigma}\overline{\mathbf{\Sigma}} = \Sigma_{ijklm}\frac{1}{2}\bigg(\Phi_{[ij|\alpha\beta}\overline{\Sigma}_{\alpha\beta |klm]}
 \\ 
-\frac{i}{5!}\epsilon_{ijklmabcde}\Phi_{ab\alpha\beta}\overline{\Sigma}_{\alpha\beta cde}\bigg).
\end{multline}
The antisymmetrization of the first product in the parentheses is on the indices $(i,j,k,l,m)$, and it is defined by
\begin{multline}
\label{antisym1}
\Phi_{[ij|\alpha\beta}\overline{\Sigma}_{\alpha\beta |klm]}=\frac{1}{10}\big( 
\Phi_{ij\alpha\beta}\overline{\Sigma}_{\alpha\beta klm} \\
-\Phi_{ik\alpha\beta}\overline{\Sigma}_{\alpha\beta jlm}
-\Phi_{il\alpha\beta}\overline{\Sigma}_{\alpha\beta kjm}
-\Phi_{im\alpha\beta}\overline{\Sigma}_{\alpha\beta klj}\\
+\Phi_{jk\alpha\beta}\overline{\Sigma}_{\alpha\beta ilm}
+\Phi_{jl\alpha\beta}\overline{\Sigma}_{\alpha\beta kim}
+\Phi_{jm\alpha\beta}\overline{\Sigma}_{\alpha\beta kli}\\
-\Phi_{kl\alpha\beta}\overline{\Sigma}_{\alpha\beta jim}
-\Phi_{km\alpha\beta}\overline{\Sigma}_{\alpha\beta jli}
+\Phi_{lm\alpha\beta}\overline{\Sigma}_{\alpha\beta jki}
\big).
\end{multline}
Finally, we have 
\begin{multline}
\label{FinalDerivative}
\frac{\partial (\mathbf{\Phi}\mathbf{\Sigma}\overline{\mathbf{\Sigma}})}{\partial \mathbf{\Sigma}}=\frac{1}{2}\bigg(\Phi_{[ij|\alpha\beta}\overline{\Sigma}_{\alpha\beta |klm]}\\
-\frac{i}{5!}\epsilon_{ijklmabcde}\Phi_{ab\alpha\beta}\overline{\Sigma}_{\alpha\beta cde}\bigg).
\end{multline}
And we can note that this is the term in the $\mathbf{\overline{126}}$ representation in the contraction between the $\mathbf{210}$ and $\mathbf{\overline{126}}$ representation, since this is totally antisymmetric in its five indices, and anti-self-dual (it can be checked explicitly using\footnote{The general expression is 
\begin{multline}
\epsilon_{i_1 \dots i_k~i_{k+1}\dots i_n} \epsilon^{i_1 \dots i_k~j_{k+1}\dots j_n}\\
= k!(n-k)!~\delta_{[ i_{k+1}}{}^{j_{k+1}} \dots \delta_{i_n ]}{}^{j_n}.
\end{multline}
} $\epsilon_{abcdeijklm}\epsilon^{abcdepqrst}=(5!)^2 \delta_{[a}^p  \delta_b^q  \delta_c^r  \delta_d^s  \delta_{e]}^t$).
So, this can be written
\begin{equation}
\frac{\partial (\mathbf{\Phi}\mathbf{\Sigma}\overline{\mathbf{\Sigma}})}{\partial \mathbf{\Sigma}}=(\mathbf{\Phi}\mathbf{\overline{\Sigma}})_{\overline{\Sigma}}.
\end{equation}

In fact, this property can also permit us to calculate this derivative in another manner, by group considerations. Indeed, we considered the singlet term built from the product $\mathbf{\Phi}\mathbf{\Sigma}\overline{\mathbf{\Sigma}}$, so from $\mathbf{210}\times\mathbf{126}\times \mathbf{\overline{126}}$, which can be written $\mathbf{126}\times (\mathbf{210}\times \mathbf{\overline{126}})$.Then, the branching rules for the second term give \cite{Slansky:1981yr}
\begin{equation}
\label{BranchingRules210-126}
\mathbf{210}\times \mathbf{\overline{126}} = \mathbf{10} + \mathbf{120} + \mathbf{\overline{126}} + \mathbf{320} + \cdots.
\end{equation}
But now, the only possibility to have a singlet term from the contraction of Eq.~(\ref{BranchingRules210-126}) with the \textbf{126} representation comes with the product $\mathbf{126}\times\mathbf{\overline{126}}$, the other terms giving a vanishing value.
%\footnote{When making a contraction of all the indices in a tensorial representation, the only non vanishing terms have to be singlet, as they are indeed invariant under all the SO(10) rotations.}
So it is possible to write this singlet term as
\begin{equation}
{\mathbf{1}}_{\mathbf{210}\times\mathbf{126}\times \mathbf{\overline{126}}} \ni \mathbf{126} \times (\mathbf{210}\times \mathbf{\overline{126}})_{\mathbf{\overline{126}}}
\end{equation}
Now, it is straightforward to take the derivative with respect to the $\mathbf{126}$, since we already simplified the term we have to consider, and it gives $(\mathbf{\Phi}\mathbf{\overline{\Sigma}})_{\overline{\Sigma}}$. We see that we could in fact compute this derivative only by considering the appropriate part in the product $\mathbf{\Phi}\mathbf{\overline{\Sigma}}$, which here have to be in the $\mathbf{\overline{126}}$ representation. So, as this representation is totally antisymmetric and anti-self-dual, it was sufficient to take the antisymmetric anti-self-dual part of this product\footnote{In a similar way we decompose a rank two tensor in its symmetric and anti-symmetric part, we can decompose a tensor in its self-dual and anti-self-dual part.}. Indeed, a tensor which is anti-self-dual or totally symmetric gives a vanishing expression when contracted with a totally antisymmetric or anti-self-dual tensor. This proves in another method the result of Eq.~(\ref{FinalDerivative}).

Thus, as the singlet term coming from such a product can be written in different manner 
\begin{multline}
\mathbf{1}_{\mathbf{210}\times\mathbf{126}\times \mathbf{\overline{126}}}
=\mathbf{210}\times (\mathbf{126}\times \mathbf{\overline{126}})_{210}
=\mathbf{\overline{126}} \\
\times (\mathbf{210}\times \mathbf{126})_{126}
=\mathbf{126} \times (\mathbf{210}\times \mathbf{\overline{126}})_{\overline{126}},
\end{multline}
we can use this method to compute all the derivative we want in a direct way.

%%%%%%%%%%%%%%%%%%%%%%%%%%%%%%%%%%%%%%%%%%%%%%%%%%%%%%%%%%%%%%%%%
\subsection{Derivative terms and associated notations}
\label{AppendixDerivatives}
 
Using the methods explained in the former section, we obtain, in addition to Eq.~(\ref{FinalDerivative})
\begin{multline}
\frac{\partial (\mathbf{\Phi}\mathbf{\Sigma}\mathbf{\overline{\Sigma}})}{\partial \mathbf{\overline{\Sigma}}}=
(\mathbf{\Phi}\mathbf{\Sigma})_{\Sigma}=
\frac{1}{2}\bigg(\Phi_{[ij|\alpha\beta}\Sigma_{\alpha\beta |klm]}\\
+\frac{i}{5!}\epsilon_{ijklmabcde}\Phi_{ab\alpha\beta}\Sigma_{\alpha\beta cde}\bigg),
\end{multline}
\begin{equation}
\frac{\partial (\mathbf{\Phi}\mathbf{\Sigma}\overline{\mathbf{\Sigma}})}{\partial \mathbf{\Phi}}=
(\mathbf{\Sigma}\mathbf{\overline{\Sigma}})_\Phi=
\Sigma_{[ij|abc}\overline{\Sigma}_{|kl]abc},
\end{equation}
and
\begin{equation}
\frac{\partial (\mathbf{\Phi}\mathbf{\Phi}\mathbf{\Phi})}{\partial \mathbf{\Phi}}=(\mathbf{\Phi}\mathbf{\Phi})_{\Phi}=3 \Phi_{[ij|ab}\Phi_{ab|kl]},
\end{equation}
where the antisymmetrization of the different products are defined as in Eq. (\ref{antisym1}), by 
\begin{multline}
\Sigma_{[ij|abc}\overline{\Sigma}_{|kl]abc}=\frac{1}{6} \big(
\Sigma_{ijabc}\overline{\Sigma}_{klabc}
-\Sigma_{ikabc}\overline{\Sigma}_{jlabc}\\
-\Sigma_{ilabc}\overline{\Sigma}_{kjabc}
+\Sigma_{jkabc}\overline{\Sigma}_{ilabc}
+\Sigma_{jlabc}\overline{\Sigma}_{kiabc}\\
-\Sigma_{klabc}\overline{\Sigma}_{jiabc}\big),
\end{multline}
and
\begin{multline}
\Phi_{[ij|ab}\Phi_{ab|kl]}=\frac{1}{3}\big(
\Phi_{ijab}\Phi_{abkl}-\Phi_{ikab}\Phi_{abjl}\\
+\Phi_{ilab}\Phi_{abjk}\big).
\end{multline}

%%%%%%%%%%%%%%%%%%%%%%%%%%%%%%%%%%%%%%%%%%%%%%%%%%%%%%%%%%%%%%%%%
%\subsection{Tensor formulation of the potential}

%%%%%%%%%%%%%%%%%%%%%%%%%%%%%%%%%%%%%%%%%%%%%%%%%%%%%%%%%%%%%%%%%
\subsection{Selection rules}
\label{AppendixSelectionRules}

For the quadratic contractions, we have
\begin{equation}
\langle \mathbf{\Phi} \rangle \langle \mathbf{\Phi} \rangle^\dagger =pp^*+aa^*+bb^*,
\end{equation}
and
\begin{equation}
\langle \mathbf{\Sigma} \rangle \langle \mathbf{\Sigma} \rangle ^\dagger=\langle \mathbf{\Sigma} \rangle \langle \overline{\mathbf{\Sigma}} \rangle   = \sigma \sigma^*,
\end{equation}
which comes from the normalization of the vectors. 

The cubic contractions give
\begin{multline}
\langle \mathbf{\Phi} \mathbf{\Phi} \mathbf{\Phi} \rangle _{\mathbf{1}}=\langle \mathbf{\Phi} \mathbf{\Phi} \rangle _{\Phi} \langle \mathbf{\Phi} \rangle ^{\text{T}} = \frac{1}{9\sqrt{2}}a^3
 + \frac{ab^2}{3\sqrt{2}} + \frac{pb^2}{2\sqrt{6}},
\end{multline}
\begin{multline}
\langle \mathbf{\Phi} \mathbf{\Phi} \rangle _{\Phi} \langle \mathbf{\Phi} \rangle ^\dagger 
= \frac{1}{9\sqrt{2}}a^2a^* + \left[ \frac{1}{6\sqrt{6}}\left(2b^*bp + b^2p^*\right) 
\right. \\ \left.
+ \frac{1}{9\sqrt{2}}\left(2b^*ba + b^2a^*\right)\right],
\end{multline}
and
\begin{multline}
\langle \mathbf{\Sigma} \overline{\mathbf{\Sigma}} \mathbf{\Phi} \rangle_{\mathbf{1}}=\langle \mathbf{\Phi} \mathbf{\Sigma} \rangle _{\Sigma} \langle \mathbf{\Sigma} \rangle^\dagger=\langle \mathbf{\Phi} \overline{\mathbf{\Sigma}} \rangle _{\overline{\Sigma}} \langle \overline{\mathbf{\Sigma}} \rangle^\dagger \\
=\langle \mathbf{\Phi} \rangle  \langle \mathbf{\Sigma} \overline{\mathbf{\Sigma}} \rangle _\Phi ^\dagger 
= \sigma \sigma^* \left( \frac{1}{10\sqrt{6}} p + \frac{1}{10\sqrt{2}} a - \frac{1}{10} b \right).
\end{multline}

Finally, the quartic contractions give
\begin{multline}
\langle \mathbf{\Phi} \mathbf{\Phi} \rangle_\Phi \langle \mathbf{\Phi} \mathbf{\Phi} \rangle_\Phi^{\text{T}}=
\frac{a^4}{164}+\frac{a^2b^2}{27} + \frac{7b^4}{648}+\frac{2ab^2p}{27\sqrt{3}}+\frac{b^2p^2}{54},
\end{multline}
\begin{multline}
\langle \mathbf{\Phi} \mathbf{\Phi} \rangle_\Phi \langle \mathbf{\Phi} \mathbf{\Phi} \rangle _\Phi ^\dagger =
\frac{(aa^*)^2}{164}  +\frac{7(bb^*)^2 }{648} \\
+ \frac{1}{162} \left({a^*}^2b^2 + a^2 {b^*}^2 + 4 aa^*bb^*\right)+\frac{bb^*pp^*}{54} \\
+\frac{1}{27\sqrt{3}}\left( abb^*p^*+a^*b^*bp \right),
\end{multline}
\begin{multline}
\langle \mathbf{\Phi} \overline{\mathbf{\Sigma}} \rangle _{\overline{\Sigma}} \langle \mathbf{\Phi} \overline{\mathbf{\Sigma}} \rangle _{\overline{\Sigma}} ^\dagger
=\langle \mathbf{\Phi} \mathbf{\Sigma} \rangle_\Sigma \langle \mathbf{\Phi} \mathbf{\Sigma} \rangle _\Sigma ^\dagger\\
= \frac{1}{600}\sigma\sigma^* (\sqrt{3}a-\sqrt{6}b+p)(\sqrt{3}a-\sqrt{6}b+p)^*\\
=\frac{1}{600}\sigma\sigma^* |\sqrt{3}a-\sqrt{6}b+p|^2 ,
\end{multline}
\begin{multline}
\langle \mathbf{\Phi} \mathbf{\Phi} \rangle_\Phi \langle \mathbf{\Sigma} \overline{\mathbf{\Sigma}} \rangle _\Phi^\dagger  
= \frac{a^2\sigma\sigma^*}{180}+\frac{b^2\sigma\sigma^*}{120}\\
-\frac{ab\sigma\sigma^*}{45\sqrt{2}}-\frac{bp\sigma\sigma^*}{30\sqrt{6}},
\end{multline}
and
\begin{equation}
\langle \mathbf{\Sigma} \overline{\mathbf{\Sigma}} \rangle _\Phi \langle \mathbf{\Sigma} \overline{\mathbf{\Sigma}} \rangle _\Phi ^\dagger
=\frac{1}{60} (\sigma\sigma^*)^2.
\end{equation}

%%%%%%%%%%%%%%%%%%%%%%%%%%%%%%%%%%%%%%%%%%%%%%%%%%%%%%%%%%%%%%%%%
\subsection{Alternative formulation}     
\label{AppendixAlternativeFormulation}

We do here the correspondence with \cite{Aulakh:2003kg,Bajc:2004xe,Cacciapaglia:2013tga}, with a tilde to denote the alternative notations. These papers take the following definitions :
\begin{multline}
W= \frac{\tilde{m}}{4!}\mathbf{\Phi}^2
+ \frac{\tilde{m_{\mathbf{\Sigma}}}}{5!}\mathbf{\Sigma}\overline{\mathbf{\Sigma}}
+ \frac{\tilde{\lambda}}{4!}\mathbf{\Phi}^3 \\
+ \frac{\tilde{\eta}}{4!}\mathbf{\Phi}\mathbf{\Sigma}\overline{\mathbf{\Sigma}}
+ \kappa S (\frac{\mathbf{\Sigma}\overline{\mathbf{\Sigma}}}{5!}-M^2),
\end{multline}
and
\begin{equation}
\left\{
\begin{array}{l}
 \tilde{p}= \Phi_{1234} ,\\
\tilde{a}= \Phi_{5678} = \Phi_{5690} = \Phi_{7890} ,\\
\tilde{b}= \Phi_{1256} = \Phi_{1278} = \Phi_{1290} \\
~~ = \Phi_{3456} = \Phi_{3478} = \Phi_{3490} ,\\
\frac{1}{\sqrt{2^5}}\tilde{\sigma} =\Sigma_{1,3,5,7,9}, \\
\frac{1}{\sqrt{2^5}} \tilde{\overline{\sigma}}= \overline{\Sigma}_{1,3,5,7,9},
\end{array}
\right.
\end{equation}
where we did not note all the possible configuration for $\mathbf{\Sigma}$ and $\mathbf{\overline{\Sigma}}$.
The link between the different definitions is
\begin{equation}
\left\{
\begin{array}{l}
\displaystyle{\frac{\tilde{m}}{4!}=\frac{m}{2}},\\[6pt]
\displaystyle{\frac{\tilde{m}_{\Sigma}}{5!}=m_{\Sigma}},\\[6pt]
\displaystyle{\frac{\tilde{\lambda}}{4!}=\frac{\lambda}{3}},
\end{array}
\right.
~~~~~~~~~
\left\{
\begin{array}{l}
\displaystyle{\frac{\tilde{\eta}}{4!}=\eta,}\\[6pt]
\displaystyle{\frac{\tilde{\kappa}}{5!}=\kappa,}\\[6pt]
\sqrt{5!}\tilde{M}=M,
\end{array}
\right.
\end{equation}
and
\begin{equation}
\left\{
\begin{array}{l}
\displaystyle{\tilde{p}=\frac{p}{\sqrt{4!}},}\\[6pt]
\displaystyle{\tilde{a}=\frac{a}{\sqrt{4!3}},}\\[6pt]
\displaystyle{\tilde{b}=\frac{b}{\sqrt{4!6}},}
\end{array}
\right.
~~~~~~~~~
\left\{
\begin{array}{l}
\displaystyle{\tilde{\sigma}=\frac{\sigma}{\sqrt{5!}}},\\[6pt]
\displaystyle{\tilde{\overline{\sigma}}=\frac{\overline{\sigma}}{\sqrt{5!}}.}
\end{array}
\right.
\end{equation}

Thus, expressing them with these particular conventions, we obtain 
\begin{multline}
W=\tilde{m}(\tilde{p}^2+3\tilde{a}^2+6\tilde{b}^2)+2\tilde{\lambda}(\tilde{a}^3+3\tilde{p}\tilde{b}^2+6\tilde{a}\tilde{b}^2)\\
+\tilde{m}_{\Sigma}\tilde{\sigma}\tilde{\overline{\sigma}}
+ \tilde{\eta}\tilde{\sigma}\overline{\sigma}(\tilde{p}+3\tilde{a}-6\tilde{b}) 
+\tilde{\kappa} s (\tilde{\sigma}\tilde{\overline{\sigma}}- \tilde{M}^2),
\end{multline}
for the superpotential, and
%\begin{widetext}
\begin{equation}
\left\{
\begin{array}{l}
\displaystyle{ F_{p}=\frac{1}{2\sqrt{6}} \left( 2\tilde{m}\tilde{p}+6\tilde{\lambda} \tilde{b}^2+\tilde{\eta} \tilde{\sigma} \tilde{\overline{\sigma}}\right)},\\
\displaystyle{ F_{a}=\frac{1}{6\sqrt{2}} \left[ 3\left(2\tilde{m}\tilde{a}+2\tilde{\lambda} (2\tilde{b}^2+\tilde{a}^2)+\tilde{\eta} \tilde{\sigma}\tilde{\overline{\sigma}}\right) \right]},\\
\displaystyle{ F_{b}=\frac{1}{12} \left[ 6\left(2\tilde{m}\tilde{b}+2\tilde{\lambda} \tilde{b}(2\tilde{a}+\tilde{p})-\tilde{\eta} \tilde{\sigma}\tilde{\overline{\sigma}}\right) \right]},\\
\displaystyle{ F_{\sigma}=\frac{1}{2\sqrt{30}}\left[ \tilde{\overline{\sigma}}\left(\tilde{m}_{\Sigma}+\tilde{\eta} (\tilde{p}+3\tilde{a}-6\tilde{b})+\tilde{\kappa} s\right)\right]},\\
\displaystyle{ F_{\overline{\sigma}}=\frac{1}{2\sqrt{30}}\left[\tilde{\sigma}\left(\tilde{m}_{\Sigma}+\tilde{\eta} (\tilde{p}+3\tilde{a}-6\tilde{b})+\tilde{\kappa} s\right)\right]},\\
\displaystyle{ F_{S}=\tilde{\kappa} (\tilde{\sigma} \tilde{\overline{\sigma}}-\tilde{M}^2)},
\end{array}
\right.
\end{equation}
for the $F$-terms. Ref.~\cite{Bajc:2004xe} and \cite{Cacciapaglia:2013tga} also introduce different $F$-terms, which are defined by, e.g. 
\begin{equation}
\tilde{F}_{\tilde{a}}=\frac{\partial{W}}{\partial\tilde{a}}.
\end{equation}
They are related to the $F$-terms by
\begin{equation}
\left\{
\begin{array}{l}
\displaystyle{ F_{p}=\frac{1}{2\sqrt{6}}  \tilde{F}_{\tilde{p}}, }\\[6pt]
\displaystyle{F_{a}=\frac{1}{6\sqrt{2}}\tilde{F}_{\tilde{a}},}\\[6pt]
\displaystyle{F_{b}=\frac{1}{12}\tilde{F}_{\tilde{b}},}
\end{array}
\right.
~~~~~~~~~
\left\{
\begin{array}{l}
\displaystyle{F_{\sigma}=\frac{1}{2\sqrt{30}}\tilde{F}_{\tilde{\sigma}},}\\[6pt]
\displaystyle{F_{\overline{\sigma}}=\frac{1}{2\sqrt{30}}\tilde{F}_{\tilde{\overline{\sigma}}},}\\[6pt]
\displaystyle{F_{S}=\tilde{F}_{\tilde{S}}.}
\end{array}
\right.
\end{equation}
Finally, as functions of these $\tilde{F}$-terms, the potential terms are
\begin{multline}
V_{\Phi}={F_{\Phi}}{F_{\Phi}}^\dagger=\frac{1}{24}\tilde{F}_{\tilde{p}}\tilde{F}_{\tilde{p}}^*+\frac{1}{72}\tilde{F}_{\tilde{a}}\tilde{F}_{\tilde{a}}^*+\frac{1}{144}\tilde{F}_{\tilde{b}}\tilde{F}_{\tilde{b}}^*,
\end{multline}
and 
\begin{equation}
V_{\Sigma}=V_{\overline{\Sigma}}={F_{\Sigma}}{F_{\Sigma}}^\dagger = \frac{1}{120} \tilde{F}_{\tilde{\sigma}}\tilde{F}_{\tilde{\sigma}}^*.
\end{equation}
At this point, we can note that the potential cannot be obtained simply by summing the square of the norms of the $\tilde{F}$ terms, but that some numerical coefficients appear. It is particularly important when comparing the values of these different potential terms, since these coefficients can considerably modify the results obtained.

%%%%%%%%%%%%%%%%%%%%%%%%%%%%%%%%%%%%%%%%%%%%%%%%%%%%%%%%%%%%%%%%%
\subsection{Set of VEV for the scheme of $G'_2$}
\label{AppendixVEVG2}

The values of the non-vanishing VEV for the SSB scheme going through $G'_2=3_C 2_L 1_R 1_{B-L}$ are
\begin{equation}
\left\{
\begin{array}{l}
\tilde{\sigma}_0=0,\\
\tilde{a}_0=-18 \sqrt{2},\\
\tilde{b}_0=\pm 18 \text{i}, \\
\tilde{p}_0= 9 \sqrt{6},\\
\end{array}
\right.
\end{equation}
before the end of inflation, and
\newpage
\begin{equation}
\left\{
\begin{array}{l}
\tilde{\sigma}_1=|1|,\\
\displaystyle{\tilde{a}_1=-18 \sqrt{2}+\frac{\left(\frac{1}{50} \pm \frac{i}{25}\right) x}{\sqrt{2}}},\\[7pt]
\displaystyle{\tilde{b}_1=\pm 18 i+\left(\frac{3}{100}\pm\frac{i}{100}\right) x},\\[7pt]
\displaystyle{\tilde{p}_1=9 \sqrt{6}-\frac{\left(\frac{1}{25}\pm\frac{9 i}{50}\right) x}{\sqrt{6}}},\\[7pt]
\displaystyle{\tilde{S}_1=-1+\frac{\alpha_1 \alpha_3}{\alpha_2}\left(\frac{4\pm 3 i}{1500}\right) \left[x+(540\pm 270 i)\right]},\\
\end{array}
\right.
\end{equation}
after the end of inflation.

%%%%%%%%%%%%%%%%%%%%%%%%%%%%%%%%%%%%%%%%%%%%%%%%%%%%%%%%%%%%%%%%%
\subsection{Characteristic radii for the toy-model string}
\label{AppendixRCharac}

Using the Lagrangian of Eq. (\ref{LagToyModel}), two characteristic radius appear for this toy-model limit. Indeed, writing the equations of motion for $f$ and $Q$ in a dimensionless form, we obtain
\begin{equation}
\displaystyle{ 2 \left( \frac{ \text{d}^2 f}{ \text{d} \tilde{r}^2} + \frac{1}{\tilde{r}} \frac{ \text{d} f} { \text{d} \tilde{r}}\right) = \frac{f
Q^2}{\tilde{r}^2} + 2 f \left( f^2-1\right) },
\end{equation}
and 
\begin{equation}
\displaystyle{\text{Tr} \left({\tau_{\text{str}}}^2 \right) \left(\frac{ \text{d}^2 Q}{ \text{d} \tilde{r}^2} - \frac{1}{\tilde{r}} \frac{ \text{d} Q}{ \text{d} \tilde{r}}\right) = \frac{2 g^2}{\kappa^2} f^2 Q}.
\end{equation}
The first equation is properly written in a dimensionless form, and as $\tilde{r} = r \kappa M$, we identify the characteristic radius of the field $f$ to be $r_f \sim {\kappa M}^{-1}$. In regard to the field $Q$, we obtain a properly dimensionless equation of motion after introducing $\rho = \tilde{r}/\kappa$, yielding 
\begin{equation}
\displaystyle{\text{Tr} \left({\tau_{\text{str}}}^2 \right)  \left(\frac{ \text{d}^2 Q}{ \text{d} \rho^2} - \frac{1}{\rho} \frac{ \text{d} Q}{ \text{d} \rho}\right) = 2 g^2 f^2 Q}.
\end{equation}
It gives the characteristic radius for $Q$ to be $r_Q \sim M^{-1}$, since $\rho = r M$.

Taking into account the previous results, we can broadly evaluate their contribution to the Lagrangian density, from Eq.~(\ref{LagEffDim}). The contribution of the field $f$ is of order $\kappa^2 M^4$, and the one of the field $Q$ of order $M^4$ (we estimate $g$ and $\text{Tr} \left({\tau_{\text{str}}}^2 \right) $ to be of order unity). So, in the limit $\kappa \geq 1$, the main contribution of the Lagrangian density comes from the field $f$, and for $\kappa\leq1$, the main contribution comes from the field $Q$. It means that in both cases, the characteristic radius of the string will be either $r_f\sim (\kappa M)^{-1}$, or $r_Q \sim M^{-1}$, see Sec.~\ref{PartDefRadius}. Note that whatever the limit we consider, the characteristic energy per unit length of the string is always of order $M^2$.

%%%%%%%%%%%%%%%%%%%%%%%%%%%%%%%%%%%%%%%%%%%%%%%%%%%%%%%%%%%%%%%%%

\bibliographystyle{unsrt}

\end{document}